\documentclass[11pt,a4paper]{article}
\usepackage{amsmath,amssymb,amsfonts,a4wide,graphicx,bm,times,url}
\usepackage{color}
\usepackage{mcite}
\usepackage{cite}
\usepackage{tikz}
\usepackage{gridset}

\makeatletter \let\old@startsection=\@startsection
\renewcommand{\@startsection}[6]
{\old@startsection{#1}{#2}{#3}{#4}{#5}{#6\mathversion{bold}}}
\makeatother

\makeatletter

\@addtoreset{equation}{section}
\makeatother
\let\refOld\ref
\renewcommand{\ref}[1]{(\refOld{#1})}

\newcommand{\superp}[2]{\genfrac{}{}{0pt}{}{#1}{#2}}

 \def\d{\delta}

 \def\p{\partial}
 
 \def\a{\alpha}
 \def\b{\beta}
 
 \def\d{\delta}
 \def\e{\epsilon}

 \def\l{\lambda}

 \def\z{\zeta }

 \def\L{\Lambda}

\def\equskip{\!\!\!\!\!\!\!\!} %Huit espaces n\UTF{00E9}gatifs
\def\la{\left\langle}
\def\ra{\right\rangle}

\def\Op{\mathcal{O}}

\def\implies{\quad\Rightarrow\quad}

\def\CZ{{\mathcal{Z}}}

\def\Zv{\mathcal{Z}_\text{vect.}}
\def\Zbf{\mathcal{Z}_\text{bfd.}}
\def\Zf{\mathcal{Z}_\text{fund.}}
\def\bZbf{\bar{\mathcal{Z}}_\text{bfd.}}

\def\CY{\mathcal{Y}}

\def\aY{|\vec{a},\vec{Y}\rangle}
\def\Ga{|G,\vec{a}\rangle}

\def\CV{\mathcal{V}}

\def\Zi{\mathcal{Z}_{\text{inst}}}
\def\nQ{Q}
\def\mf{m^{(f)}}

\newcommand{\ba}{\begin{eqnarray}}
\newcommand{\ea}{\end{eqnarray}}

\begin{document}
% \vspace*{-2cm}
% \begin{flushright}
% \jobname .pdf\\ \today
% \end{flushright}
\begin{titlepage}
\renewcommand{\thefootnote}{\fnsymbol{footnote}}
\vspace*{-2cm}
\begin{flushright}
% \jobname .pdf\\ \today\\
 UT-15-41
\end{flushright}

\vspace*{1cm}
    \begin{Large}
%    \begin{bf}
       \begin{center}
%         {\huge SH$^c$ algebra on quiver partition functions and qq-characters}
	{\huge Holomorphic field realization of SH$^c$}\\
	{\huge and quantum geometry of quiver gauge theories}
       \end{center}
%    \end{bf}   
    \end{Large}
\vspace{0.7cm}

\begin{center}
{\Large Jean-Emile Bourgine$^\dagger$, Yutaka Matsuo$^\ast$, Hong Zhang$^\diamond$}
\\[.4cm]
{\em {}$^\dagger$INFN Bologna, Universit\`a di Bologna}\\
{\em Via Irnerio 46, 40126 Bologna, Italy}
\\[.4cm]
{\em {}$^\ast$ Department of Physics, The University of Tokyo}\\
{\em Bunkyo-ku, Tokyo, Japan}
\\[.4cm]
{\em {}$^\diamond$ Department of Physics and Center for Quantum Spacetime (CQUeST)}\\
{\em Sogang University, Seoul 04107, Korea}\\[.4cm]
\texttt{ bourgine\,@\,bo.infn.it,\quad  matsuo\,@\,phys.s.u-tokyo.ac.jp,\quad kilar\,@\,sogang.ac.kr}
\end{center}

\vspace{0.7cm}

\begin{abstract}
\noindent
In the context of 4D/2D dualities, SH$^c$ algebra, 
introduced by Schiffmann and Vasserot, provides a 
systematic method to analyse the instanton partition functions of
$\mathcal{N}=2$ supersymmetric gauge theories.
In this paper, we rewrite the SH$^c$ algebra in terms of three 
holomorphic fields $D_0(z)$, $D_{\pm1}(z)$
with which the algebra and its representations are
simplified. 
%Unlike the conventional 2D CFT,
%the parameter $z$ is not interpreted as the coordinate
%of the Riemann surface but the spectral parameter
%in the integrable models.
The instanton partition functions for 
arbitrary $\mathcal{N}=2$ super Yang-Mills theories
with $A_n$ and $A^{(1)}_n$ type quiver
diagrams are compactly expressed as a product of 
four building blocks,
Gaiotto state, dilatation, flavor vertex operator and intertwiner
which are written in terms of SH$^c$ and the orthogonal basis 
introduced by Alba, Fateev, Litvinov and Tarnopolskiy.
These building blocks are characterized by new
conditions
which generalize the known ones on the Gaiotto
state and the Carlsson-Okounkov vertex.
Consistency conditions of the inner product give algebraic
relations for the chiral ring generating functions defined
by Nekrasov, Pestun and Shatashvili. 
In particular we show
the polynomiality of the qq-characters which have been introduced 
as a deformation of the Yangian characters. 
These relations define a second quantization of the Seiberg-Witten geometry, 
and, accordingly, reduce to a Baxter TQ-equation in the Nekrasov-Shatashvili 
limit of the Omega-background. 

%and find that they provides a simpler description of
%Gaiotto states. We also
%The action of SH$^c$ algebra on quiver gauge theories is reformulated in terms of a bosonic free field with positive modes. The generators of degree zero and plus/minus one that span the whole algebra are written as holomorphic series $D_0(z)$, $D_{\pm1}(z)$. The diagonal series $D_0(z)$ plays the role of the current of the bosonic field. The remaining series $D_{\pm1}(z)$ can be interpreted as creation/anihilation operators, their action on Gaiotto states is worked out in terms of vertex operators. Their adjoint action on the intertwining operator, necessary to describe bifundamental multiplets, takes a similar form. 

%These identities are further used to show the polynomiality of the qq-characters recently introduced by Nekrasov, Pestun and Shatashvili  They 
%A new characterization of the building block for quiver $\mathcal{N}=2$ instanton partition functions under the action of SH$^c$ generators is derived. It is used to generate the qq-character formulas introduced by Nekrasov, Pestun and Shatashvili.
\vspace{0.5cm}
%\noindent\textbf{Keywords:} Mayer Expansion, Thermodynamical Bethe Ansatz, $\mathcal{N}=2$ SUSY Gauge Theory, Instantons\\
\end{abstract}

\vfill

\end{titlepage}
\vfil\eject

\setcounter{footnote}{0}

\section{Introduction}
SH$^c$ is an algebra introduced by Shiffmann and Vasserot in \cite{schiffmann2013cherednik} (see also \cite{arbesfeld2012presentation}) to describe the equivariant cohomology of the instanton moduli space of $\mathcal{N}=2$ gauge theories in four dimensions. It has been defined as a spherical (symmetric) version of the degenerate double affine Hecke algebra (DAHA) which has been developed by Cherednick for many years \cite{cherednik2005double}.\footnote{In fact, SH$^c$ is a short notation introduced in \cite{schiffmann2013cherednik} for \underline{c}entral extension of the \underline{S}pherical degenerate double affine \underline{H}ecke algebra. It may be better referred to as Schiffmann-Vasserot algebra, but we will use SH$^c$ in the text.} While DAHA encodes the algebraic (recursive) properties of Macdonald polynomials \cite{macdonald1998symmetric} with two deformation parameters, degenerate DAHA is obtained by taking a limit $q,t\to1$ such that one parameter $\beta$ remains, with $t=q^{-\b}$. In this limit, Macdonald polynomials degenerate into Jack polynomials.

This algebra precisely describes the algebraic structure behind Nekrasov instanton partition functions \cite{Nekrasov2004} with the Omega background $\mathbb{R}_{\e_1}^2\times\mathbb{R}_{\e_2}^2$ with the identification $\beta=-\e_1/\e_2$ and has been used to prove the 4D/2D correspondence which generalizes Alday, Gaiotto and Tachikawa's proposal \cite{Alday2010} (AGT conjecture) for various types of quiver gauge theories -- namely pure super Yang-Mills theory \cite{schiffmann2013cherednik}, the gauge theories with fundamental \cite{Matsuo2014}
and bifundamental hypermultiplets \cite{Kanno2013} (see also
the recent preprint \cite{neguct2015exts}). For pure super Yang-Mills, the theory is characterized by a Gaiotto state \cite{Gaiotto:2009ma}, a coherent state affiliated to the Whittaker vector appearing in the representation theory of noncompact Lie algebra. To address higher quiver gauge theories, an operator which intertwines different representations is required for the description of bifundamental multiplets. 
%In the simplest case, namely the case of (adjoint) matter $U(1)$ gauge theory, such an operator has been defined by Carlsson and Okounkov in \cite{carlsson2008exts}. 
It is defined by a direct product of the Carlsson-Okounkov operator
\cite{carlsson2008exts} which describes the $U(1)$ part and
the vertex operator of Toda field theory.
This operator must be properly generalized to describe the gauge theories on arbitrary quiver diagrams. % which contains gauge groups with different ranks.
%\footnote{Recently a geometrical description (Ext operator) of bifundamental multiplets appeared in } where results similar to \cite{Kanno2013} were found.}
 Aside from SUSY gauge theories, SH$^c$ has also revealed itself particularly useful in the study of vortex dynamics \cite{fujimori20152d}.

In such developments, the important role devoted to SH$^c$ as a ``universal" symmetry came principally from the fact that it contains all $W_N$ algebras for arbitrary $N$, together with an additional $U(1)$ factor. The parameter $\beta$ is identified with a deformation parameter that defines the central charge $c=(N-1)(1+Q^2N(N+1))$ of $W_N$ representations through the combination $Q=\sqrt\beta-\sqrt\beta^{-1}$. In this sense, SH$^c$ should be regarded as a one parameter deformation of the $W_{1+\infty}$ algebra. The latter is known to have realizations in terms of $N$ free fermions acting on a space of $N$-tuple Young diagrams. These representations, that we call here rank $N$ representations, are identical to those defined by Fateev and Lukyanov in \cite{Fateev:1987zh}. The correspondence between the two algebras has also been confirmed in more general cases
where the Hilbert space contains singular vectors. The most typical example is the minimal models of $W_N$. There, it has been demonstrated explicitly in \cite{fukuda2015sh} that SH$^c$ reproduces the proper descriptions of the Hilbert space constrained by the so-called N-Burge conditions
\cite{bershtein2014agt,alkalaev2014conformal}. This universality is essential when we have to treat a system that contains gauge groups of different rank, as it is the case for quiver theories.

On the other hand, the action of SH$^c$ on instanton partition functions of quiver $\mathcal{N}=2$ gauge theories is very different from the representation of $W_N$ algebras. It is better understood after the introduction of an orthonormal basis constructed by Alba, Fateev, Litvinov and Tarnopolsky (AFLT basis) to prove the 4D/2D duality \cite{Alba2011,Fateev:2011hq}. AFLT basis should be regarded as a generalization of Jack polynomials \cite{stanley1989some, Estienne:2011qk, Morozov:2013rma,Smirnov2014}, it is the proper basis to describe the action of degenerate DAHA. Instead of the description in terms of chiral primary fields with different spins, it is defined more abstractly through generators $D_{m,n}$ with two indices $m,n$. The first index $m\in \mathbb{Z}$ is identified with the index of the Virasoro generators $L_m$ while the second one $n\in \mathbb{Z}^{\geq 0}$ corresponds to the spin $n+1$ of the generator. Since it is a nonlinear symmetry with a reasonably complicated structure, we have to be careful how to organize the generators. One of the authors \cite{Bourgine2014a} has recently found that holomorphic expansion in terms of the {\em second} index, $D_{\pm 1}(z), D_0(z)$, gives a compact description of SH$^c$ through the study of the Nekrasov-Shatashvili \cite{nekrasov2009quantization} limit of AGT conjecture. This turns out to be very useful and is a main tool of this paper.

The focus of the paper is to provide an SH$^c$ description of Nekrasov partition functions for general $A_Q$ and $A^{(1)}_Q$-type quiver gauge theories. To do so, the action of SH$^c$ operators on Gaiotto states, together with the adjoint action on the intertwiner operator describing bifundamental fields, is worked out. 
%These actions are conveniently expressed in terms of the vertex operators associated to a free boson with only positive modes. 
These actions are conveniently expressed in terms of the vertex operators $\CY(z)$ associated with the current $D_0(z)$.
They extend the work on the covariance of the partition function presented in \cite{Kanno2013} by giving us the possibility to consider quiver theories with gauge groups of different ranks. As a consequence of our results, several useful identities can be established among correlators of the gauge theories. In particular, we were able to recover the expression of the qq-characters recently introduced by Nekrasov, Pestun and Shatashvili (NPS) \cite{Nekrasov2015}. For the simplest $A_1$ case with
fundamental multiplets \ref{equ_qq},
$$
\chi(z)=\la \CY(z+\e_+)+q\dfrac{m(z)}{\CY(z)}\ra.
$$
Here $\e_+:=\e_1+\e_2$ and $\la\cdots\ra$ denotes an average weighted by the instanton partition function, which is defined in \ref{def_avr}. 
The operator $\CY(z)$ is interpreted as an operator version
of the chiral ring generating function.
These characters, presented as further deformation of the characters of Yangian algebras \cite{Knight1995}, encode in a compact form a recursion relation among the instanton partition functions \cite{Nekrasov2013}. 
%They are defined as operators with the property to be polynomial when considered in an average between two Gaiotto states (or more generally as a sum over Young diagrams with the extra contribution defining the instanton partitions, see the definitions \ref{def_avr} and \ref{A2_average}). 
Here we show that SH$^c$ provides a proper symmetry behind the qq-character formulae, as was already predicted by NPS, and that the polynomiality property naturally follows from our description.

The qq-characters define a double deformation of the Seiberg-Witten geometry in a form of second quantization. In the above example, the Seiberg-Witten curve is expressed as \ref{SW_curve}, $$y+q\dfrac{m(z)}{y}=\prod_{\ell=1}^N(z-a_\ell).$$
Seiberg-Witten theory is well-known to provide an effective description of the infrared sector of $\mathcal{N}=2$ gauge theories on $\mathbb{R}^4$ \cite{Seiberg1994a,Seiberg1994}. The effective Lagrangian is written in terms of an holomorphic function, the prepotential, obtained from the knowledge of an algebraic curve and a differential form. This formulation is identified with the construction of finite gap solutions for classical integrable hierarchies, the algebraic curve corresponding to the spectral curve of the system \cite{Marshakov1999}.\footnote{These finite gap solutions can also be obtained from Hitchin systems.} When the gauge theory is considered in the Nekrasov-Shatashvili (NS) limit $\e_2\to0$ of the Omega-background, the associated integrable systems are quantized, with the remaining parameter $\e_1$ playing the role of the Planck constant \cite{nekrasov2009quantization}. The algebraic curve becomes the Baxter TQ-equation of the quantum system \cite{Poghossian2010,Fucito2011} (see also \cite{Fucito2012,Nekrasov2013} for the extension to quivers), it is equivalent to a Schr\"odinger equation under a quantum change of variables \cite{Bourgine2012a}, in a form of ODE/IM correspondence \cite{Dorey2007}.\footnote{In this classical version of AGT correspondence, the Shr\"odinger equation is obtained as the semiclassical limit of the null vector decoupling equations obeyed by Liouville correlators containing a degenerate operators. It is sometimes referred as the \textit{bispectral duality} \cite{Mironov2012b,Mironov2012a}.} In this framework, the two complex variables of the algebraic curve become non-commutative, thus defining a first quantization of the Seiberg-Witten curve \cite{Mironov2009b,Mironov2010a,Mironov2010}. In the full Omega-background, the qq-character is an operator acting in a Hilbert space of quantum states. In the NS limit, the expectation value of this operator in the Gaiotto state (which plays the role of a coherent state) becomes the T-polynomial of the TQ-equation, while its defining relation in terms of vertex operators reproduces Baxter's relation. In this sense, the qq-character formula presents a second quantization of the integrable system in which the TQ-relation emerges in the classical $\e_2\to0$ limit.

% Recently Nekrasov, Pestun and Shatashvili found a compact expression of the recursion relation among the instanton partition functions \cite{Nekrasov2013} in a more general set-up. In particular, a quantum version of character formulae (so called qq-character) plays an essential role. In this work, Nekrasov and others also presented a higher rank generalization of their character formulae. We show that SH$^c$ provides a proper symmetry behind their formulae, as they have already predicted. We need some technical development though. The Gaiotto state needs more precise description than it was considered and the intertwining operators should describe the gauge groups with different ranks. The latter property can be covered by the symmetry which does not depend on the rank $N$ of $W_N$ which characterizes SH$^c$. We find that the holomorphic description of
% the algebra provides such generalizations in a compact form. The recursion formula in \cite{Nekrasov2013} can be reproduced through a simple algebra among holomorphic generators which resembles that of 2D CFT.  

We organize the paper as follows. In section 2, we introduce the holomorphic field description of SH$^c$ algebra and the rank $N$ representation. We also provide useful expressions for the adjoint actions of the vertex operators. In section 3, after a brief review of the general construction of the instanton partition functions, we introduce the building blocks (Gaiotto states, flavor vertex operator, intertwiner) with which the partition functions are written as a product. We show that the Gaiotto state satisfies stronger constraints which are compactly expressed in terms of SH$^c$ fields.  A generalized intertwiner which connects different rank gauge groups is also defined. It satisfies similar conditions as the Gaiotto states and indeed it reduces to the Gaiotto state when the gauge group of one side is trivial. The flavor vertex operator is used to include the fundamental hypermultiplets in the gauge theories. These results are used in section 4 to build an infinite number of constraints among the correlation functions of the vertex operator $\CY$. The new characterizations of the Gaiotto states and the intertwiner play an essential role to give a closed and compact expression for these constaints -- written in the form of qq-characters.  Finally in section 5 we present their interpretation as quantum Seiberg-Witten geometry along the line of \cite{Nekrasov:2012xe, Nekrasov2013, Bourgine2014a}. The concluding section proposes some perspectives for future research, and several technical details are gathered in the appendix.

% 
% {\color{red} Possible story of the paper:
% \begin{enumerate}
% \item Rewriting SH$^c$ algebra in terms of holomorphic fields
% \item algebra between them and vertex operators
% \item Writing every Nekrasov partition function for A-type quiver in
% terms of Gaiotto state and intertwiner
% \item Evaluation of $D_{\pm 1}(z)$ on Gaiotto state and intertwiner
% \item Derivation of NPS identities from SH$^c$
% \item Interpretation of qq-character from SH$^c$
% \end{enumerate}
% }

\section{Reformulation of SH$^c$ algebra}
\subsection{SH$^c$ algebra in terms of holomorphic fields}
The SH$^c$ algebra is defined on a set of operators $D_{m,n}$ with the double grading $(m,n)\in\mathbb{Z}\times\mathbb{Z}^{\geq0}$ \cite{schiffmann2013cherednik}. The first index is called the \textit{degree} and the second index the \textit{order}. 
The algebraic relations involving $D_{\pm 1, n}$ and $D_{0,n}$ are written as
\ba
\left[D_{0,n} , D_{\pm 1,m} \right] & =&\pm D_{\pm1,n+m-1}, \;\;\; n \geq 1 \,,\label{SH1}\\
\left[D_{-1,n},D_{1,m}\right]&=&E_{n+m} \;\;\; n,m \geq 0\label{SH3}\,,\\
\left[D_{0,n} , D_{0,m} \right] & =& 0 \,,\,\, n,m\geq 0\,.\label{SH4}
\ea
where $E_k$ denotes a linear combination of powers of the generators 
$D_{0,n}$ that will be given shortly. Additional relations can be found in \cite{arbesfeld2012presentation}, but they will not be used here. The algebra is spanned by the operators of degree $0$ and $\pm1$ 
upon the recursive use of the following commutation relations,
\begin{equation}\label{def_hg}
D_{\pm(m+1),0}=\pm\dfrac1m[D_{\pm1,1},D_{\pm m,0}],\quad D_{\pm m,n}=\pm[D_{0,n+1},D_{\pm m,0}],
\end{equation}
for $n\geq0$ and $m>0$.
Rank $N$ representations of SH$^c$ match with those of a
semidirect product of the $W_N$ algebra and a U(1) current. 
The Heisenberg generators ($U(1)$
currents) are related to $D_{m,0}$, the Virasoro generators to $D_{m,0},D_{m,1}$, 
and operators $D_{m,n}$ with $n>1$ to the currents of spin $n+1$
\cite{schiffmann2013cherednik, Matsuo2014}.

It is useful to assemble the generators in the form of holomorphic fields\cite{Bourgine2014a},
\begin{equation}
D_{\pm1}(z)=\sum_{n=0}^\infty{z^{-n-1}D_{\pm1,n}},\quad D_0(z)=\sum_{n=0}^\infty{z^{-n-1} D_{0,n+1}},\quad E(z)=1+\e_+\sum_{n=0}^\infty{z^{-n-1} E_n},
\end{equation}
where $\e_+=\e_1+\e_2$. We use here the Omega-background equivariant deformation parameters $\e_1, \e_2$ \cite{Nekrasov2004, Nekrasov2003a} instead of the SH$^c$ deformation parameter $\beta=-\e_1/\e_2$ in order to simplify the comparison with gauge theories.\footnote{The correspondence between the convention of \cite{Kanno2013} and this paper is summarized in the appendix \refOld{app:conv}.}
It is noted that these Laurent series are vanishing at $z=\infty$, 
in which they are different from usual holomorphic fields in CFT. 

We rewrite the defining properties of the generators $D_{\pm1,n}$ and $D_{0,n}$
in terms of the holomorphic fields,
%\footnote{These relations are invariant under the rescaling $D_{0,n+1}\to \a^n D_{0,n+1}$ and $D_{\pm1,n}\to \a^n D_{\pm1, n}$. This property has been used here with $\a=\e_2$ to simplify the definition of the generating series.}
\begin{equation}\label{comm_SHc}
[D_0(z),D_{\pm1}(w)]=\pm\dfrac{D_{\pm1}(w)-D_{\pm1}(z)}{z-w},\quad [D_{-1}(z),D_1(w)]=\dfrac{E(w)-E(z)}{z-w}\e_+^{-1}.
\end{equation}
For the definition of $E(z)$ and the vertex operators which will appear later, 
it is essential to introduce
\begin{equation}\label{Phi}
\Phi(z):=\log(z) D_{0,1}-\sum_{n=1}^\infty 
\frac{1}{n z^n}D_{0,n+1}\implies D_0(z)=\p_z\Phi(z).
\end{equation}
This definition of the field $\Phi(z)$ resembles the mode expansion of an holomorphic free bosonic field in CFT and
the series $D_0(z)$ can be interpreted as the associated current. 
As we noted, however,
the usual U(1) current in CFT is expanded as a sum over the degree
as $J(\zeta) =\sum_{n\in\mathbb{Z}} \mathrm{(coeff.)} D_{-n,0} 
\zeta^{-n-1}$ while \ref{Phi} is expanded with respect to the order. In addition, fields at different points are commuting, $[\Phi(z), \Phi(w)]=0$, as a consequence of \ref{SH4}. In this sense, the interpretation of the complex variable $z$ is different from the holomorphic coordinate of a Riemann surface but it should rather be seen as the spectral parameter of an integrable model. We will come back to this description later.
%\footnote{
%SH$^c$ was originally defined which gives the recursive structure
%among Jack polynomials and $D_{0,n}$ are interpreted as
%the commuting charges of Calogero-Sutherland model.
%This is particularly true for $U(1)$ gauge theories for which the operator $D_{0,2}$ is identified with the Calogero-Moser Hamiltonian while the generators $D_{0,n}$ with $n>2$ give the higher conserved quantities. For higher rank gauge groups, the Hopf algebra structure of SH$^c$ can be exploited to build tensorial versions of these quantities \cite{Smirnov2013}.
%}
%
%By analogy, it is appealing to regroup the $D_{0,n}$ generators in the expansion of a ``scalar field'' with only positive (commuting) modes.

The following dressed combination of vertex operators will play a central role in our reformulation of the SH$^c$ algebra and the correspondence with gauge theories,
\begin{equation}\label{def_CY}
\CY(z)=e^{c(z)} e^{\Phi(z-\e_1)}e^{\Phi(z-\e_2)}e^{-\Phi(z)}e^{-\Phi(z-\e_+)}.
\end{equation}
The function $c(z)$ encodes the dependence in the infinite
number of the central charges $c_n$ ($n\geq 0$) of the algebra. It expands as
\begin{eqnarray}
c(z)= c_0\log(z)-\sum_{n=1}^\infty \frac{c_n}{n z^n}.
\end{eqnarray}
%As a matter of fact, 
The generating series $E(z)$ can now be expressed using the newly defined vertex operator,
\begin{equation}\label{E_CY}
E(z)=\CY(z+\e_+)\CY(z)^{-1}.
\end{equation}

\subsection{Rank $N$ representations}
Among possible representations of SH$^c$, the best studied one is the rank $N$ representation where the Hilbert space is spanned by a basis labeled by $N$-tuple Young diagrams $\vec Y=(Y_1,\cdots, Y_N)$. The representation is characterized by $N$ complex numbers $a_\ell$ ($\ell=1,\cdots, N$) that define the central charges $c_n$ through the relation%\footnote{In order to simplify the equations, the coefficients $a_\ell$ have been rescaled by a factor $-\e_2$ and then shifted by $\e_+$. They are related to the standard Coulomb branch vevs $\tilde{a}_\ell$ of the gauge theory by $a_\ell=-\e_2\tilde{a}_\ell+\e_+$.}
\begin{eqnarray}
e^{c(z)}=\prod_{\ell=1}^{N}(z-a_\ell).%\quad \mathcal{C}(z)=\prod_{\ell=1}^{N} \frac{z-a_\ell-\xi}{z-a_\ell}\,.
\end{eqnarray}
%In SH$^c$, the central charges specifies the representations
To emphasize the dependence of the representation in the parameters $a_\ell$ through the central charges, they will be included in the notation of the vector basis $|\vec{a}, \vec{Y}\rangle$. These vectors form an orthonormal basis of the representation space,
\begin{equation}
\langle \vec{a}, \vec{Y}|\vec{a}, \vec{Y'}\rangle=\d_{\vec Y,\vec Y'},\quad 1=\sum_{\vec{Y}} |\vec{a},\vec{Y}\rangle \langle\vec{a},\vec{Y}|.
\end{equation}
For $N=2$, this basis is actually proportional to the one employed in the proof of AGT conjecture in \cite{Alba2011} and is usually referred as the AFLT basis. For a generic $N$, they can be identified with the generalized Jack polynomials introduced in \cite{Morozov:2013rma} and studied in \cite{Smirnov2014}. The vacuum state is obtained by taking the $N$-tuple of empty Young diagrams denoted $\vec{\emptyset}$, it satisfies
\begin{equation}
D_{0,n}|\vec{a},\vec{\emptyset}\rangle=D_{-1,n}|\vec{a},\vec{\emptyset}\rangle=0,\quad\text{or}\quad D_0(z)|\vec{a},\vec{\emptyset}\rangle=D_{-1}(z)|\vec{a},\vec{\emptyset}\rangle=0.
\end{equation}
The Hilbert space spanned by $|\vec{a},\vec{Y}\rangle$ will be denoted $\mathcal{V}_{\vec a}$. When several Hilbert spaces are considered, an extra label $\vec a$ will be inserted on the notation of the operators $D^{\vec a}_{r}(z)$ to specify in which space $\mathcal{V}_{\vec a}$ they act.
The rank $N$ representations of SH$^c$ are equivalent to the representations of $\mathcal{W}_N\times U(1)$ \cite{schiffmann2013cherednik, fukuda2015sh}.

\begin{figure}\label{fig:AYRY}
\begin{center}
\begin{tikzpicture}[scale=0.7]
\draw (2,-2) node {$Y$};
%\filldraw [thin, fill=yellow!20] (5,0) rectangle (6,-1);
\filldraw [fill=yellow!30] (5,0) rectangle (5.5,-0.5);
\filldraw [fill=yellow!30] (3,-2) rectangle (3.5,-2.5);
\filldraw [fill=yellow!30] (1,-4) rectangle (1.5,-4.5);
\filldraw [fill=yellow!30] (0,-5) rectangle (0.5,-5.5);
\filldraw [fill=red!20] (4.5,-1.5) rectangle (5,-2);
\filldraw [fill=red!20] (2.5,-3.5) rectangle (3,-4);
\filldraw [fill=red!20] (0.5,-4.5) rectangle (1,-5);
\draw [very thick] (0,0) -- (5,0);
\draw [very thick] (5,0) -- (5,-2);
\draw [very thick] (5,-2) -- (3,-2);
\draw [very thick] (3,-2) -- (3,-4);
\draw [very thick] (3,-4) -- (1,-4);
\draw [very thick] (1,-4) -- (1,-5);
\draw [very thick] (1,-5) -- (0,-5);
\draw [very thick] (0,0) -- (0,-5);

\filldraw [fill=yellow!20] (9,-2) rectangle (9.5,-2.5);
\draw (11,-2.3) node {$ \in A(Y)$};
\filldraw [fill=red!20] (9,-4) rectangle (9.5,-4.5);
\draw (11,-4.3) node {$ \in R(Y)$};
\end{tikzpicture}
\end{center}
\caption{$A(Y)$ and $R(Y)$}
\end{figure}
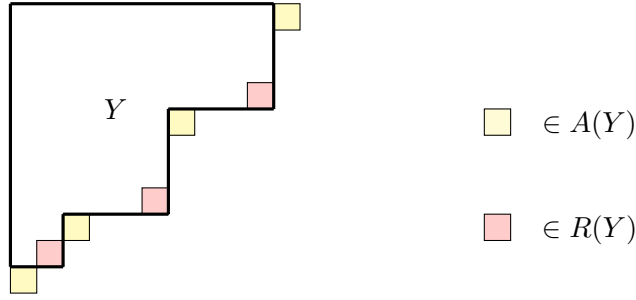

The action of SH$^c$ generators of degrees $\pm1$ on the state $\aY$ involves the $N$-tuple Young diagram with a box added/removed. As such, they can be seen as 
an analog of 
creation/annihilation operators while the total number of boxes 
in $\vec Y$ represents the number of particles (later identified with the instanton charge). Following \cite{Bourgine2014a}, $N$-tuple Young diagrams $\vec{Y}$ with a box $x$ added/removed will be denoted $\vec{Y}\pm x$ (respectively). 
We further introduce the sets $A(\vec{Y})$ and $R(\vec{Y})$ containing all the boxes that can be added to/removed from the Young diagrams composing $\vec{Y}$. 
In the figure \refOld{fig:AYRY}, we illustrate the
locations of the boxes in the sets $A(Y)$ and $R(Y)$ with the example of a single Young diagram $Y$.
The boxes $x\in\vec{Y}$ are characterized by a triplet of indices $(\ell,i,j)$ where $\ell=1\cdots N$ and $(i,j)\in Y_\ell$ gives the position of the box in the $\ell$th Young diagram. To each box $x$ is associated a complex number $\phi_x$ depending on the central charges using the map
\begin{equation}
x=(\ell,i,j)\in \vec{Y}\quad\longrightarrow\quad \phi_x=a_\ell+(i-1)\e_1+(j-1)\e_2\in\mathbb{C}.
\end{equation}
With these definitions, the action of the spanning subalgebra takes the simple form \cite{Kanno2013,Bourgine2014a}
\begin{eqnarray}
&& D_{+1}(z)\aY=\sum_{x\in A(\vec{Y})}\dfrac{\L_x(\vec{Y})}{z-\phi_x}|\vec{a},\vec{Y}+ x\rangle,\quad 
D_{-1}(z)\aY=\sum_{x\in R(\vec{Y})}\dfrac{\L_x(\vec{Y})}{z-\phi_x}|\vec{a},\vec{Y}- x\rangle,\nonumber\\
&& D_0(z)\aY=\sum_{x\in \vec{Y}}\dfrac1{z-\phi_x}\aY,\label{act_D}
\end{eqnarray}
which are equivalent to their component form ($n\geq 0$):
\begin{eqnarray}
&&D_{+1,n}\aY=\sum_{x\in A(\vec{Y})}
(\phi_x)^n\L_x(\vec{Y})|\vec{a},\vec{Y}+x\rangle,\quad
D_{-1,n}\aY=\sum_{x\in R(\vec{Y})}
(\phi_x)^n\L_x(\vec{Y})|\vec{a},\vec{Y}-x\rangle,
\nonumber\\
&& D_{0,n+1}\aY=\sum_{x\in \vec{Y}}(\phi_x)^n\aY.\label{act_D1}
\end{eqnarray}
We note that the second relation in \ref{act_D1} implies that the moments of $\phi_{x\in \vec Y}$ are the eigenvalues of the commuting charges $D_{0,n}$. In the (generalized) Calogero-Sutherland system, $D_{0,n}$ plays the role of infinite commuting charges and $\phi_x$ is interpreted as the momentum of each particle.  The interpretation of $z$ as the spectral parameter is natural in this sense. The left action of SH$^c$ generators on bra $\langle\vec a,\vec Y|$ is identical for the diagonal operators $D_0(z)$, $E(z)$, $e^{\Phi(z)}$. However it is reversed for the operators $D_{\pm1}(z)$,
\begin{equation}\label{act_D_adj}
\langle\vec a,\vec Y|D_{+1}(z)=\sum_{x\in R(\vec{Y})}\dfrac{\L_x(\vec{Y})}{z-\phi_x}\langle\vec a,\vec Y- x|,
\quad
\langle\vec a,\vec Y|D_{-1}(z)=\sum_{x\in A(\vec{Y})}\dfrac{\L_x(\vec{Y})}{z-\phi_x}\langle\vec a,\vec Y+ x|.
\end{equation}
The series $E(z)$ is also diagonal on the states $\aY$, with eigenvalues given by the function 
\begin{equation}\label{L2}
\L(z)^2=\prod_{x\in A(\vec{Y})}\dfrac{z-\phi_x+\e_+}{z-\phi_x}\prod_{x\in R(\vec{Y})}\dfrac{z-\phi_x-\e_+}{z-\phi_x}.
\end{equation}
The coefficients $\L_x(\vec{Y})$ in the action \ref{act_D} of $D_\eta(z)$ correspond to the residues of this function $\L(z)^2$ at $z=\phi_x$ with $x\in A(\vec Y)\mbox{ or }R(\vec Y)$:
\begin{equation}\label{prop_L}
\L(z)^2=1+\e_+\sum_{x\in A(\vec{Y})}\dfrac{\L_x(\vec{Y})^2}{z-\phi_x}-\e_+\sum_{x\in R(\vec{Y})}\dfrac{\L_x(\vec{Y})^2}{z-\phi_x},\quad \L_x(\vec{Y})^2=\prod_{\superp{y\in A(\vec{Y})}{y\neq x}}\dfrac{\phi_x-\phi_y+\e_+}{\phi_x-\phi_y}\prod_{\superp{y\in R(\vec{Y})}{y\neq x}}\dfrac{\phi_x-\phi_y-\e_+}{\phi_x-\phi_y}.
\end{equation}% The coefficients $\L_x(\vec Y)$ have a sign ambiguity that it fixed as follows: \textcolor{red}{How???}.
Eventually, the action of the vertex operator is expressed in terms of a product over the boxes of $\vec Y$,
\begin{equation}\label{Vertex_aY}
e^{\Phi(z)}\aY=Q_{\vec{Y}}(z)\aY,\quad\text{with}\quad Q_{\vec{Y}}(z)=\prod_{x\in\vec{Y}}(z-\phi_x).
\end{equation}
The specific combination of vertex operators entering in the definition \ref{def_CY} of $\CY(z)$ leads to a remarkable simplification of its eigenvalues
\begin{equation}\label{prop_CY}
\CY(z)\aY=\prod_{\ell=1}^{N}(z-a_\ell) 
\prod_{x\in \vec{Y}}\frac{(z-\phi_x-\e_1)(z-\phi_x-\e_2)}{(z-\phi_x)(z-\phi_x-\e_+)}\aY
=
\dfrac{\prod_{x\in A(\vec{Y})}(z-\phi_x)}{\prod_{x\in R(\vec{Y})}(z-\e_+-\phi_x)}\aY.
\end{equation}
We note that there is a cancellation of factors between the numerators
and the denominators in the middle term, and the resulting expression in the RHS bears contributions only from the edges of the Young diagrams.
Taking the ratio \ref{E_CY} defining the operator $E(z)$, we recover the expression \ref{L2} for the function $\L(z)^2$:
\begin{equation}
\label{Ez}
E(z)\aY=\L(z)^2\aY\,.
\end{equation}
In the appendix \refOld{sec:commutator},
we provide an explicit computation of the commutation relations
of $D_0(z), D_{\pm 1}(z)$ in the rank $N$ representation.

Finally we would like to mention the existence of an automorphism of representation.
Under the shift of $\vec{a}$, $a_i\rightarrow \vec{a}'=\vec{a}+\mu\vec e$ where $\vec e=(1,1,\cdots, 1)$, the representation \ref{act_D1} implies that
\begin{eqnarray}
D^{\vec a+\mu \vec e}_{+1}(z) |\vec{a}+\mu\vec e,\vec Y\rangle&=&\sum_{x\in A(\vec{Y})}\dfrac{\L_x(\vec{Y})}{z-\mu-\phi_x}|\vec{a}+\mu \vec e,\vec{Y}+ x\rangle,\\
D^{\vec a+\mu \vec e}_{-1}(z) |\vec{a}+\mu\vec e,\vec Y\rangle&=&\sum_{x\in R(\vec{Y})}\dfrac{\L_x(\vec{Y})}{z-\mu-\phi_x}|\vec{a}+\mu \vec e,\vec{Y}- x\rangle\,.
\end{eqnarray}
The coefficients appearing here may be identified with the representation of $D^{\vec a}_{\pm1}(z-\mu)$.
It implies that there is an automorphism of the algebra by shifting the variable $z$:
$D^{\vec a+\mu \vec e}_{r}(z)\sim D^{\vec a}_{r}(z-\mu)$ for $r=0,\pm 1$.
This shift symmetry of the representations is referred as the spectral flow
in the context of $W_{1+\infty}$-algebra\cite{frenkel1995central}.

\subsection{Adjoint action of the vertex operators}
In order to prepare for the computations necessary in the next sections, we would like to evaluate the commutation relations 
between the vertex operators $e^{\pm\Phi(z)}$ 
and the elements spanning the SH$^c$ algebra. 
The generators of degree zero form a commutative subalgebra, 
as a consequence the field $\Phi(z)$ commute with the series $D_0(z)$. 

The evaluation of the adjoint action on $D_{\pm 1}$ is slightly more involved.
We introduce a vertex operator depending on two finite sets of points $z_i$ and $w_j$ with $i\in I$, $j\in J$,
\ba
U(\{z_i\},\{w_j\}):=\exp\left(\sum_{i\in I}\Phi(z_i)-\sum_{j\in J} \Phi(w_j)\right).
\ea
We claim the following identities:
\begin{align}
\begin{split}
\label{UDU3}
U(\{z_i\},\{w_j\})^{-1}D_{1}(u) U(\{z_i\},\{w_j\})&=\mathrm{P}^-_{u=\infty,z_{i\in I}}\left[\frac{\prod_{j\in J} (w_j-u)}{\prod_{i\in I}(z_i-u)} D_{1}(u)\right]\,,\\
U(\{z_i\},\{w_j\})^{-1}D_{-1}(u) U(\{z_i\},\{w_j\})&=\mathrm{P}^-_{u=\infty,w_{j\in J}}\left[\frac{\prod_{i\in I} (z_j-u)}{\prod_{j\in J}(w_j-u)} D_{-1}(u)\right]\,,
\end{split}
\end{align}
with the projector $\mathrm{P}^-_{u=\infty,z_{i\in I}}$ 
acting on functions of the variable $u$ as
\begin{eqnarray}\label{def_proj}
\mathrm{P}^-_{u=\infty,z_{i\in I}} f(u):=f(u)-\sum_{i\in I} \frac{\mbox{Res}_{\zeta=z_i}f(\zeta)}{u-z_i} -\mathrm{P}_u^+f(u).
\end{eqnarray}
Here $\mathrm{P}^+_z$ picks up the positive powers of a Laurent series in $z$,
namely for a function $f(z)=\sum_{n=-m}^\infty a_n z^{-n}$, 
it operates as $\mathrm{P}^+_zf(z)=\sum_{n=-m}^0 a_n z^{-n}$.
Later we will use a similar notation for the orthogonal projector
$\mathrm{P}^-_z=1-\mathrm{P}^+_z=\mathrm{P}^-_{z=\infty}$
which picks up the negative powers of $f(z)$.
The second term in \ref{def_proj} also plays the role to remove singularities at $u=z_i$.
One may use a contour integration to write these projections in a compact form,
for example,
\begin{equation}
\mathrm{P}^-_{u=\infty,z_{i\in I}} f(u)=\oint_C{\dfrac{f(w)}{u-w}\dfrac{dw}{2\pi i}},
\end{equation}
where the contour $C$ is defined by $|w|=R$ with
$R<\mbox{Min}_i(|z_i|)$.  The formula \ref{UDU3} formally resembles an
OPE in CFT, up to the existence of the
 projection operator which is necessary here
since there is no singularity except for $u=0$ on the left hand side.

%the simplest examples of the adjoint action of the vertex operators on the elements $D_{\pm1}(z)$ can be deduced.\footnote{
%We demonstrate the derivation of these formulae
%for more general cases later.}
%%\footnote{In fact, the vertex operator $e^{\Phi(z)}$ is the exponential of an element of the Lie algebra, as such it can be seen as a Lie group element. We are describing here the adjoint representation of a group element on the Lie algebra.}
%\begin{align}
%\begin{split}\label{V_Deta}
%e^{-\eta\Phi(z)}D_{\eta}(w)e^{\eta\Phi(z)}&=\frac{D_{\eta}(w)-D_{\eta}(z)}{z-w},\\%=\mathrm{P}^-_{w=z}\left[\frac{D_{\eta}(w)}{z-w}\right],\\
%e^{\eta\Phi(z)}D_{\eta}(w)e^{-\eta\Phi(z)}&=(z-w)D_{\eta}(w)+D_{\eta,0}=\mathrm{P}^-_{w}\left[(z-w)D_{\eta}(w)\right],
%\end{split}
%\end{align}
%where $\eta=\pm$ is a sign.  
%In the following, we sometimes use the notation $\mathrm{P}^-_z=\mathrm{P}^-_{z=\infty}$ and $\mathrm{P}^+_z=1-\mathrm{P}^-_{z=\infty}$ for the orthogonal projections on regular/singular parts at infinity. For instance, 
%It is interesting to note that the RHS in the first line of \ref{V_Deta} reproduces the commutation relation between $D_0(z)$ and $D_{\eta}(z)$ given in \ref{comm_SHc} (up to a sign $\eta$). 

Before proving \ref{UDU3}, it may be instructive
to give some specific examples which will be used later.
The first one is when one of the sets $I,J$ is null:
\begin{align}
\begin{split}\label{com_D_Phi}
e^{-\eta\sum_{i=1}^M \Phi(z_i)}D_{\eta}(z_0)e^{\eta\sum_{i=1}^M \Phi(z_i)}&=\sum_{i=0}^M D_{\eta}(z_i)\prod_{j(\neq i)} \frac{1}{z_j-z_i},\\
e^{\eta\sum_{i=1}^M\Phi(z_i)}D_{\eta}(w)e^{-\eta\sum_{i=1}^M\Phi(z_i)}&=\mathrm{P}^-_w\left[D_{\eta}(w)\prod_{i=1}^{M}(z_i-w)\right].
\end{split}
\end{align}
The other one is the adjoint action of $\CY(z)$
which is defined as a product of vertex operators with shifted arguments: 
%its adjoint action on SH$^c$ operators is a special case of the general formula \ref{UDU3},
\begin{align}
\begin{split}
&\dfrac1{\CY(z)}D_{-1}(w)\CY(z)=S(w-z)D_{-1}(w)+\dfrac{\e_1\e_2}{\e_+}\left(\dfrac{D_{-1}(z)}{z-w}-\dfrac{D_{-1}(z-\e_+)}{z-w-\e_+}\right),\\
&\CY(z+\e_+)D_{1}(w)\dfrac1{\CY(z+\e_+)}=S(z-w)D_{1}(w)-\dfrac{\e_1\e_2}{\e_+}\left(\dfrac{D_{1}(z)}{z-w}-\dfrac{D_{1}(z+\e_+)}{z-w+\e_+}\right),
\end{split}
\end{align}
where $S(z)$ denotes a scattering factor $S(z)$ defined as
\begin{equation}\label{braiding}
S(z)=\dfrac{(z+\e_1)(z+\e_2)}{z(z+\e_+)}.
\end{equation}

\paragraph{Proof of the formula \ref{UDU3}}
To end up this section, we would like to give a short derivation of the identity \ref{UDU3}. 
Rather than working with the commutator \ref{comm_SHc}
directly, it is easier to evaluate the action on
the states $\aY$
which form a faithful representation of the SH$^c$ algebra.
%and the commutator between the vertex operators and the series 
%$D_{\pm1}(z)$ may be most easily evaluated through 
%the action \ref{act_D} and \ref{Vertex_aY} of the operators on these states. 
We use the property 
\begin{equation}\label{Qvar}
\dfrac{Q_{\vec{Y}\pm x}(z)}{Q_{\vec{Y}}(z)}=(z-\phi_x)^{\pm 1},
\end{equation}
which is a direct consequence of \ref{Vertex_aY}, and the action \ref{act_D} of $D_\eta(z)$ on the states $\aY$.
It follows that 
\begin{eqnarray}\label{UDU}
U(\{z_i\},\{w_j\})^{-1}D_{1}(u) U(\{z_i\},\{w_j\})|\vec{a},\vec{Y}\rangle&=&
\sum_{x\in A(\vec Y)}\frac{\prod_{j\in J} (w_j-\phi_x)}{\prod_{i\in I}(z_i-\phi_x)}
\frac{\Lambda_x(\vec Y)}{u-\phi_x}|\vec{a},\vec{Y}+x\rangle.
\end{eqnarray}
The product in the RHS can be rewritten as a sum over single poles in $\phi_x$, with an extra polynomial term, using the algebraic identity,
{\small
\begin{eqnarray}
\frac{\prod_{j\in J} (w_j-\phi)}{(u-\phi)\prod_{i\in I}(z_i-\phi)}
=\sum_{n=0}^{|J|-|I|-1} a_n(u|z,w) \phi^n +
\frac{\prod_{j\in J} (w_j-u)}{(u-\phi)\prod_{i\in I}(z_i-u)}
+\sum_{i\in I}
\frac{\prod_{j\in J} (w_j-z_i)}{(z_i-\phi)(u-z_i)\prod_{j\in I\setminus\{i\}}(z_j-z_i)}\,.
\end{eqnarray}
}
Here $a_n$ are the coefficients appearing in Laurent expansion of the LHS in $\phi$, they depend on the parameter $z_i$, $w_j$ and $u$:
\begin{equation}
\mathrm{P}_\phi^+\frac{\prod_{j\in J} (w_j-\phi)}{(u-\phi)\prod_{i\in I}(z_i-\phi)}
=\sum_{n=0}^{|J|-|I|-1} a_n(u|z,w) \phi^n.
\end{equation}
The sum over single poles in $\phi$ can be used to reform $D_1(z)$, while the polynomial part in $\phi_x$ gives the transformations $D_{1,n}$, and \ref{UDU} becomes
\begin{eqnarray}
\left(\frac{\prod_{j\in J} (w_j-u)}{\prod_{i\in I}(z_i-u)}D_1(u)
+\sum_{n=0}^{|J|-|I|-1} a_n(u|z,w) D_{1,n} 
+\sum_{i\in I}
\frac{\prod_{j\in J} (w_j-z_i)}{(u-z_i)\prod_{j\in I\setminus\{i\}}(z_j-z_i)}D_1(z_i)
\right)|\vec{a},\vec{Y}\rangle\,.
\end{eqnarray}
Since the states $\aY$ generate a faithful representation, the equality of vectors can be lifted at the level of operators
{\small
\begin{eqnarray}\label{UDU2}
U(z,w)^{-1}D_{1}(u) U(z,w)=
\frac{\prod_{j\in J} (w_j-u)}{\prod_{i\in I}(z_i-u)}D_1(u)
+\equskip\sum_{n=0}^{|J|-|I|-1} a_n(u|z,w) D_{1,n} 
+\sum_{i\in I}
\frac{\prod_{j\in J} (w_j-z_i)}{(u-z_i)\prod_{j\in I\setminus\{i\}}(z_j-z_i)}D_1(z_i)
\end{eqnarray}
}
The expression for $U(z,w)^{-1}D_{-1}(u) U(z,w)$ is similarly obtained and is written as \ref{UDU2} with the substitution of the variables $z_i\leftrightarrow w_j$. The right hand side of \ref{UDU2} can be simplified by analyzing the first term: the second term cancels the poles (and the constant part) at $u=\infty$ of the first term, while the third term cancels the simple poles at $u=z_i$.  The existence of such terms is natural since the left hand side of \ref{UDU2} is not singular at these points. The procedure of removing the unwanted poles is performed by the projector $\mathrm{P}^-_{u=\infty,z_{i\in I}}$ defined in \ref{def_proj}, and \ref{UDU2} produces \ref{UDU3}. It is noted that to analyze the pole at infinity, the following property should be employed, 
\begin{equation}
\mathrm{P}_\phi^+\dfrac{r(\phi)}{u-\phi}=\dfrac{\mathrm{P}_\phi^+r(\phi)-\mathrm{P}_u^+r(u)}{u-\phi}=-\mathrm{P}_u^+\dfrac{r(u)}{u-\phi},
\end{equation}
for any meromorphic function $r(z)$. It implies in particular
\begin{equation}
\sum_{n=0}^{|J|-|I|-1} a_n(u|z,w) \phi^n=-\sum_{n=0}^{|J|-|I|-1} a_n(\phi|z,w) u^n.
\end{equation}

\section{Instanton partition function and SH$^c$ algebra}
\subsection{Nekrasov instanton partition function}
\begin{figure}[bpt]
\begin{center}
\begin{tikzpicture}[scale=0.8]
\filldraw [fill=yellow!30] (0,0) circle (1cm);
\filldraw [fill=yellow!30] (4,0) circle (1cm);
\filldraw [fill=yellow!30] (12,0) circle (1cm);
\filldraw [fill=green!30] (-0.7,2) rectangle (0.7,3.4);
\filldraw [fill=green!30] (4-0.7,2) rectangle (4+0.7,3.4);
\filldraw [fill=green!30] (12-0.7,2) rectangle (12+0.7,3.4);
\draw [->] (1.2 ,0) -- (2.8,0);
\draw [->] (5.2 ,0) -- (6.8,0);
\draw [very thick,dashed] (7.2,0) -- (8.8,0);
\draw [->] (9.2 ,0) -- (10.8,0);
\draw (0,1) -- (0,2);
\draw (4,1) -- (4,2);
\draw (12,1) -- (12,2);
\draw (0,0) node {$SU(N_1)$};
\draw (0,-1.4) node {vevs $\vec{a_1}$};
\draw (4,-1.4) node {vevs $\vec{a_2}$};
\draw (12,-1.4) node {vevs $\vec{a_Q}$};
\draw (4,0) node {$SU(N_2)$};
\draw (12,0) node {$SU(N_Q)$};
\draw (2,0.5) node {bf. $m_{12}$};
\draw (6,0.5) node {bf. $m_{23}$};
\draw (10,0.5) node {$m_{Q-1,Q}$};
\draw (0,2.7) node {$\tilde{N}_1$};
\draw (4,2.7) node {$\tilde{N}_2$};
\draw (12,2.7) node {$\tilde{N}_Q$};
\draw (0,3.8) node {fund. $\vec m_1$};
\draw (4,3.8) node {fund. $\vec m_2$};
\draw (12,3.8) node {fund. $\vec m_Q$};
\end{tikzpicture}
\caption{$A_Q$ linear quiver}
\label{f:quiver}
\end{center}
\end{figure}
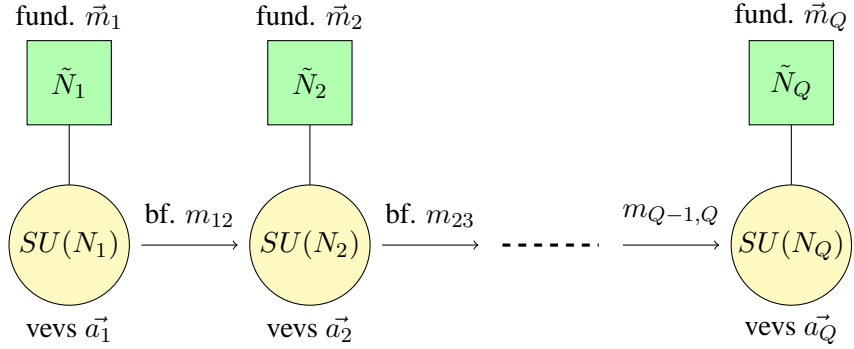

Class $\mathcal{S}$ gauge theories with $\mathcal{N}=2$ supersymmetry are obtained by compactification of the six dimensional $\mathcal{N}=(2,0)$ theory on a Riemann surface. They are classified by a quiver diagram where each node $i$ is in correspondence with the simple group component $SU(N_i)$ of the total gauge group $G=\otimes_iSU(N_i)$. Thus, to each node corresponds a gauge multiplet containing a vector, two fermions and a scalar field in the adjoint representation. The arrows $i\to j$ of the quiver represents bifundamental matter fields, i.e. a chiral multiplet containing a fermion and a scalar field, with mass $m_{ij}$, and transforming in the fundamental representation of $SU(N_i)\times SU(N_j)$. In addition, a number $\tilde{N}_i$ of fundamental (or anti-fundamental) matter fields can be attached to each node $i$. They consist in chiral multiplets of masses $m_i^{(f)}$ with $f=1\cdots \tilde{N}_i$, encoded in the $\tilde{N}_i$-vector $\vec m_i$ (see Figure \refOld{f:quiver}).

\begin{figure}[bpt]
\begin{center}
\begin{tikzpicture}[scale=0.7]
\draw (0,0) -- (10,0);
\draw (10,0) -- (10,-3);
\draw (10,-3) -- (7,-3);
\draw (7,-3) -- (7,-5);
\draw (7,-5) -- (4,-5);
\draw (4,-5) -- (4,-7);
\draw (4,-7) -- (0,-7);
\draw (0,0) -- (0,-7);
\filldraw [fill=yellow!20] (0,-1.5) rectangle (10,-2);
\filldraw [fill=red!10] (4.5,0) rectangle (5,-5);
\filldraw [fill=green!30] (4.5,-1.5) rectangle (5,-2);
\draw (4.8,0.4) node {$i$};
\draw (-0.5,-1.8) node {$j$};
\draw (10.5,-1.8) node {$\l'_j$};
\draw (4.8,-5.4) node {$\l_i$};
%\draw (4.8,-1.8) node {$x$};
\end{tikzpicture}
\end{center}
\caption{Young diagram}
\label{f:coord}
\end{figure}
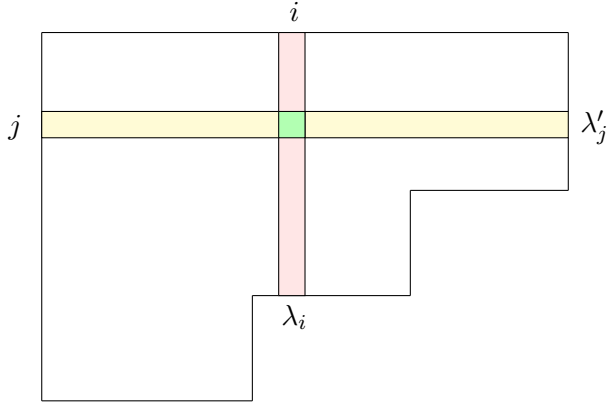

The instanton partition functions of class $\mathcal{S}$ theories have been evaluated using localization in the Omega-background \cite{Nekrasov2004}. The theory is considered on the Coulomb branch where the adjoint scalar fields take non-zero vacuum expectation values. These complex parameters will be denoted $a_\ell^{(i)}$ with $\ell=1\cdots N_i$, they form the $N_i$-vector $\vec{a}_i$ attached to the node $i$. Localization provides a sum over nested integrals that can be computed by residues. The residues are in one-to-one correspondence with the boxes of the $N_i$-tuple Young diagrams for each node $i$ of the quiver. The resulting formula is a sum over realizations of these diagrams weighted by the multiplets contributions \cite{Nekrasov2003a, losev1998issues,moore2000integrating,flume2003algorithm, bruzzo2003multi}:
\begin{equation}\label{Zinst}
\CZ_\text{inst.}=\sum_{\vec{Y}_1,\cdots\vec{Y}_{\nQ}}\prod_{i=1}^{\nQ} q_i^{|\vec Y_i|} \Zv(\vec a_i,\vec Y_i)\Zf(\vec a_i,\vec Y_i;\vec m_i) \prod_{i\to j\in E_Q} \Zbf(\vec a_i,\vec Y_i;\vec a_j,\vec Y_j|m_{ij}),
\end{equation}
where $\nQ$ is the number of nodes in the quiver, $E_Q$ its set of links, and $|\vec{Y}|$ denotes the total number of boxes in the N-tuple Young diagram $\vec{Y}$. The instanton counting parameter $q_i$ corresponds to the exponentiated gauge coupling at the node $i$, suitably renormalized in asymptotically free theories .
%%%%%%%%%%%%%%%%%%%%%%%%%%%%%%%%%%%%%%%%%%%%%%%%%%%%%
It is known that the contribution from each representation can be systematically derived from that for the  bifundamental representation. Taking a bifundamental field of mass $m_{12}$ coupled to the two gauge groups $SU(N_1)$ and $SU(N_2)$, the contribution reads:
%\textcolor{red}{There was a conflict of notations for $Y_i$ representing both the Young diagram and the height of the columns.}
\begin{eqnarray}\label{Zbfd}
\Zbf(\vec{a},\vec Y;\vec{b},\vec{W}|m_{12})
&=& \prod_{\ell=1}^{N_1} \prod_{\ell'=1}^{N_2} g_{Y_\ell,W_{\ell'}}(a_\ell-b_{\ell'}-m_{12})\\
g_{\l,\mu}(x)&=& \prod_{(i,j)\in \l}(x+\e_1(\l_j'-i+1)-\e_2(\mu_i-j))\nonumber\\
&&\cdot\prod_{(i,j)\in \mu}(-x+\e_1(\mu_j'-i)-\e_2(\l_i-j+1))\,.
\end{eqnarray}
Here $\l_i$ is the height of $i^\mathrm{th}$ column and $\l'_i$ is the length of $i^\mathrm{th}$ row of Young diagram $\l$ (see Figure \refOld{f:coord}).
 %%%%%%%%%%%%%%%%%%%%%%%%%%%%%%%%%%%%%%%%%%%%%%%%%%%%%
%%%%%%%%%%%%%%%%%%%%%%%%%%%%%%%%%%%%%%%%%%%%%%%%%%%%%
\vspace{0.5cm}

The other building blocks can be written from \ref{Zbfd} as follows.
\begin{itemize}
\item Fundamental hypermultiplets transforming under the gauge group $SU(N)$ and the flavor group $SU(\tilde{N})$, with masses $m_1,\cdots, m_{\tilde{N}}$: we take a vanishing bifundamental mass $m_{12}=0$, for the first node $N_1=\tilde{N}$, $\vec{a}_1=\vec{m}:=(m^{(1)},\cdots, m^{(\tilde{N})})$ and $\vec Y_1=\vec \emptyset$, and for the second node $N_2=N$, $\vec{a}_2=\vec a$ and $\vec Y_2=\vec Y$ arbitrary:\footnote{We have chosen to shift the definition of the fundamental masses by $\e_+$ in order to simplify formulas: $\vec m_\text{fund.}=\vec m+\e_+$, $\vec m_\text{af.}=\vec m+\e_+$. Note also that antifundamental contributions will be not discussed here as they are equivalent to fundamental contributions with shifted masses.}
\begin{eqnarray}\label{def_Zf}
% \CZ_\mathrm{fund.}(\vec{m},\vec{a},\vec{Y})&=&\prod_{q=1}^{N_f} \CZ_\mathrm{fund.}(m_q,\vec{a},\vec{Y})=\prod_{q=1}^{N_f} \prod_{x\in \vec{Y}}
% (\phi_x-m_q)=Z_\mathrm{bf}(\vec{m},\vec \emptyset;\vec{a},\vec{Y}|0)
\CZ_\mathrm{fund.}(\vec{m};\vec{a},\vec{Y})&=&\Zbf(\vec{m},\vec \emptyset;\vec{a},\vec{Y}|0).
\end{eqnarray}
\item Antifundamental hypermultiplet: in a symmetric way, we take $m_{12}=0$, for the first node $N_1=N$, $\vec{a}_1=\vec a$ and $\vec Y_1=\vec Y$ and for the second one $N_2=\tilde{N}$, $\vec{a}_2=-\vec{m}$ and $\vec Y_2=\vec \emptyset$,
\begin{eqnarray}\label{def_Zaf}
\CZ_\mathrm{af.}(\vec{m};\vec{a},\vec{Y})&=&\Zbf(\vec{a},\vec{Y};-\vec{m},\vec \emptyset|0)=\CZ_\mathrm{fund.}(-\e_+-\vec m;\vec{a},\vec{Y}).
%\CZ_\mathrm{af.}(\vec{m};\vec{a},\vec{Y})&=&\prod_{p=1}^{N_f} \CZ_\mathrm{af.}(m_p,\vec{a},\vec{Y})=\prod_{p=1}^{N_f} \prod_{x\in \vec{Y}}(\phi_x-\e_++m_q)=\CZ_\mathrm{bf}(\vec{a},\vec{Y};-\vec{m},\vec \emptyset|0)
\end{eqnarray}
%We note that $\CZ_\mathrm{af.}(\vec m,\vec{a},\vec{Y})=\CZ_\mathrm{fund.}(\e_+-m,\vec{a},\vec{Y})$.
\item Adjoint hypermultiplet: we take $N_1=N_2=N$, $\vec Y_1=\vec Y_2=\vec Y$ and $\vec a_1=\vec a_2=\vec a$,
\begin{eqnarray}
\CZ_\mathrm{adj.}(\vec{a},\vec{Y}|m)&:=& \Zbf(\vec{a},\vec{Y};\vec{a},\vec{Y}|m).
\end{eqnarray}
\item Vector multiplet: inverse of the adjoint hypermultiplet with zero mass,
\begin{eqnarray}
\Zv(\vec{a},\vec{Y})&:=& \Zbf(\vec{a},\vec{Y};\vec{a},\vec{Y}|0)^{-1}.
\end{eqnarray}
\end{itemize}

We note that the fundamental matter can be seen as bifundamental matter
 where one of the two gauge groups is taken in the weak coupling limit, effectively becoming a flavor group. 
 In this limit, the corresponding exponentiated gauge coupling $q$ is sent to zero, and due to the presence of the factor $q^{|\vec{Y}|}$, only empty Young diagrams contribute in the summations. 
 As a result, the contribution of a fundamental matter multiplet is derived from the bifundamental contribution by attaching to each set of $\tilde{N}_i$ fundamental flavors an $\tilde{N}_i$-tuple of empty Young diagrams.

The fact that the various contributions to the partition function of the different multiplets are derived from the bifundamental contribution  implies important consequences for their SH$^c$ realization that will be presented in the next section.
%%%%%%%%%%%%%%%%%%%%%%%%%%%%%

%The vector contribution is obtained from this expression as
%\begin{equation}
%\Zv(\vec a,\vec Y)=\dfrac1{\Zbf(\vec a,\vec Y;\vec a,\vec Y|0)}.
%\end{equation}
%\begin{equation}\label{def_Zf}
%\Zf(\vec m;\vec a,\vec Y)=\Zbf(\vec m,\vec\emptyset;\vec a,\vec Y|0).
%\end{equation}

\subsection{Action of SH$^c$ operators on instanton partition functions}
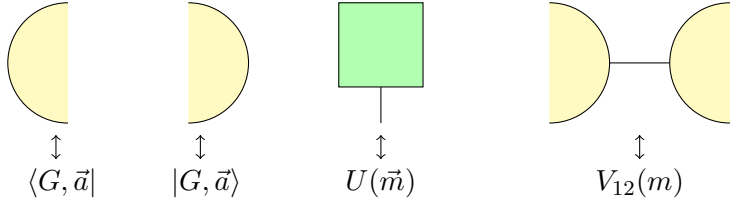
\begin{figure}
\begin{center}
\begin{tikzpicture}[scale=0.8]
\filldraw[fill=yellow!30] (0,1) arc (90:270:1);
\draw (-0.1,-2) node {$\langle G,\vec a|$};
\draw [<->] (-0.2,-1.2) -- (-0.2,-1.6);
\filldraw[fill=yellow!30] (2,-1) arc (270:450:1);
\draw [<->] (2.2,-1.2) -- (2.2,-1.6);
\draw (2.3,-2) node {$| G,\vec a\rangle $};
\filldraw [fill=green!30] (4.5,-0.4) rectangle (5.9,1);
\draw (5.2,-0.4) -- (5.2,-1);
\draw [<->] (5.2,-1.2) -- (5.2,-1.6);
\draw (5.2,-2) node {$U(\vec m) $};
\filldraw[fill=yellow!30] (8,-1) arc (270:450:1);
\draw (9,0) -- (10,0);
\filldraw[fill=yellow!30] (11,1) arc (90:270:1);
\draw [<->] (9.5,-1.2) -- (9.5,-1.6);
\draw (9.5,-2) node {$V_{12}(m) $};
\end{tikzpicture}
\end{center}
\caption{Correspondence between quiver diagram and Gaiotto states/
vertex operator/intertwiner}
\label{f:qS}
\end{figure}

In this paper we focus on the linear quiver $A_{\nQ}$ and its affine version $A_{\nQ}^{(1)}$, they are characterized by the set of arrows $E_Q=\{i\to i+1,i=1\cdots \nQ-1\}$ and $E_Q=\{i\to i+1,i=1\cdots \nQ\}$ respectively, with the identification of indices modulo $\nQ$. One of the goal of this paper is to formulate the action of SH$^c$ operators on the (affine) linear quiver instanton partition function. For this purpose, we need to rewrite the partition function in terms of elements of the representation theory of SH$^c$: Gaiotto states, intertwiner 
and the vertex operator (Figure \refOld{f:qS}).

\paragraph{Gaiotto state}
To each node $i$ of the quiver diagram is associated a vector space of representation $\CV_{\vec{a_i}}$ spanned by the vectors $|\vec{a}_i,\vec Y_i\rangle$ where $\vec{Y}_i$ takes values in all the possible realization of $N_i$-tuple Young diagrams. The set of complex parameters $\vec{a}_i$ is fixed in each $\CV_{\vec{a_i}}$ and define the central charges of the representation of SH$^c$. The Gaiotto state has been introduced in \cite{Gaiotto:2009ma} as a specific Whittaker vector of the Virasoro algebra with respect to the maximal nilpotent subalgebra $\{L_n, n>0\}$. This algebra is spanned by the two elements $L_1$ and $L_2$, and the Gaiotto state is defined up to a normalization by the conditions,\footnote{As a consequence of the Virasoro commutation relations, the second condition implies $L_n|G\rangle=0$ for $n>2$.}
\begin{equation}\label{cond_Gaiotto}
L_1|G\rangle=\a |G\rangle,\quad L_2|G\rangle=0,
\end{equation}
where $\a$ is a constant. This definition has been generalized to the case with fundamental flavors \cite{Marshakov:2009gn}, and to higher rank \cite{Kanno:2011fw,Kanno:2012xt}, and eventually implemented in the space of representation of SH$^c$ (which contains a Virasoro sub-algebra) \cite{schiffmann2013cherednik},
\begin{equation}\label{def_Gaiotto}
|G,\vec a\rangle=\sum_{\vec Y}\sqrt{\CZ_\text{vect}(\vec a,\vec{Y})}|\vec a, \vec{Y}\rangle,\quad |G,\vec a\rangle\in\CV_{\vec{a}}.
\end{equation}
This state is known to provide the instanton partition function of pure $\mathcal{N}=2$ SYM ($A_1$ quiver, $\tilde{N}=0$),
\begin{eqnarray}\label{Zinst_A1_pure}
\mathcal{Z}_\text{inst}=\langle G,\vec a|q^{D}|G,\vec a\rangle =\sum_{\vec Y}q^{|\vec Y|}\Zv(\vec a, \vec Y),
\end{eqnarray}
where the operator $D=D_{0,1}$ counts the number of boxes in $\vec Y$, $D\aY= |\vec{Y}|\aY$. It is identified with $L_0$ in Virasoro algebra
up to the zero mode and $q^D$ may be regarded as the propagator
in string theory.
In the following, we refer to the operator of the form $q^D$
as the dilatation operator. It satisfies,
\begin{eqnarray}\label{dilatation}
q^D D_{\pm 1}(z) = D_{\pm 1}(z) q^{D\pm 1}\,.
\end{eqnarray}

In terms of SH$^c$ operators written in the form of holomorphic fields,
the Gaiotto state has a new characterization.
This is one of the main results of the paper:
\begin{eqnarray}\label{action_Gaiotto}
D_{-1}(z)|G,\vec{a}\rangle&=&
\dfrac1{\sqrt{-\e_1\e_2}}\dfrac1{\CY(z)}|G,\vec{a}\rangle,\\
D_{1}(z)|G,\vec{a}\rangle&=&\dfrac{-1}{\sqrt{-\e_1\e_2}}
\mathrm{P}^-_z\CY(z+\e_+)|G,\vec{a}\rangle\label{action_Gaiotto2},\\
\langle G,\vec a|D_{-1}(z)&=&\dfrac{-1}{\sqrt{-\e_1\e_2}}\langle G,\vec a|\mathrm{P}^-_z\CY(z+\e_+),\label{action_Gaiotto_adj}\\
\langle G,\vec a|D_{1}(z)&=&\dfrac1{\sqrt{-\e_1\e_2}}\langle G,\vec a|\dfrac1{\CY(z)}\,.
\label{action_Gaiotto_adj2}
\end{eqnarray}
These formulae are a consequence of a more general result, presented in \ref{action_vertex1} and \ref{action_vertex2} below, and proven in appendix \refOld{AppA}.
% a new characterization of the Gaiotto state in terms of action of the generating series $D_\eta(z)$ is derived,
%\footnote{From \ref{act_D_adj} it is easy to deduce that the adjoint action on bra $\langle G,\vec a|$ is reversed,
%	\begin{equation}
%	\end{equation}}

These new expressions contain more information than the previously known relations given in \ref{prev_Gaiotto}. They reveal themselves powerful enough to derive several useful relations, presented in \cite{Nekrasov2013}, among instanton partition function for arbitrary (A-type) quiver diagrams. The asymptotic of the operators $\CY(z)$ at infinity is deduced from \ref{prop_CY}: $\CY(z)\sim z^N$ since $|A(\vec Y)|-|R(\vec Y)|=N$ for any $N$-tuple $\vec Y$. Expanding the first relation at infinite spectral parameter $z$ allows to recover the characterization of the Gaiotto states in \cite{Kanno2013,Matsuo2014},
\begin{equation}\label{prev_Gaiotto}
D_{-1,n}\Ga=0,\quad D_{-1,N-1}\Ga=\dfrac1{\sqrt{-\e_1\e_2}}\Ga,\quad D_{-1,N}\Ga=\dfrac1{\sqrt{-\e_1\e_2}}\left(\sum_{\ell=1}^N a_\ell\right)\Ga,
\end{equation}
where $n=1\cdots N-2$, and the last property has been obtained using the formula (A.3) in \cite{Bourgine2014a}. These identities suggest to see the Gaiotto state as a (partial) coherent state in the physical sense of eigenstate of the annihilation operators $D_{-1,n}$.
% It is known to satisfy the condition for Whittaker vector:
% \begin{eqnarray}\label{Gaiotto}
% W^{(k)}_r |G,\vec a\rangle =(\mbox{constant})|G,\vec a\rangle
% \end{eqnarray}
% for some $k\leq n$ and $r>1$ for rank $N$ representation. They are derived through similar condition in terms of SH$^c$ generators:
% \begin{eqnarray}\label{Gaiotto2}
% D_{-l,d}|G,\vec a\rangle=\mathrm{(const)}|G,\vec a\rangle\quad (d=1,\cdots, N). 
% \end{eqnarray}

\paragraph{Flavor vertex operator} Due to the presence of the empty Young diagram in the definition \ref{def_Zf}, the fundamental matter contribution can be written in a simpler form,
\begin{equation}
\Zf(\vec m;\vec a,\vec Y)=\prod_{x\in \vec Y}\prod_{f=1}^{\tilde{N}}(\phi_x-m^{(f)})=\prod_{f=1}^{\tilde{N}}(-1)^{|\vec Y|}Q_{\vec Y}(m^{(f)}),
\end{equation}
where $Q_{\vec Y}(z)$ denotes the eigenvalue of the vertex operator defined in \ref{Vertex_aY}. 
This expression implies that the vertex operator
can be used to insert fundamental multiplets
in the quiver gauge theories.
% for the introduction of the mass operator $U(\vec m):\CV_{\vec a}\to \CV_{\vec a}$ defined as
\begin{equation}
U(\vec m)=(-1)^{\tilde{N} D}\exp\left(\sum_{f=1}^{\tilde{N}}\Phi(m^{(f)})\right)\implies U(\vec m)\aY=\Zf(\vec m;\vec a,\vec Y)\aY.
\end{equation}
This operator generates the modified Gaiotto states in the presence of fundamental multiplets, as studied in \cite{Matsuo2014}. 
Since it plays the role to add the contribution of fundamental hypermultiplets with flavor group $SU(\tilde{N})$, it will sometimes be referred to as the flavor vertex operator.
The instanton partition function for this theory can be written
\begin{equation}\label{Zinst_A1}
\mathcal{Z}_\text{inst}=\langle G,\vec a|q^{D}U(\vec m)|G,\vec a\rangle =\sum_{\vec Y}q^{|\vec Y|}\Zv(\vec a, \vec Y)\Zf(\vec m;\vec a,\vec Y).
\end{equation}
It is noted that the vertex operator $U(\vec m)$ commutes with %the instanton number operator $D$.
the dilatation operator $q^D$.

\paragraph{Intertwiner} Up to now, only partition functions of $\mathcal{N}=2$ theories with a single gauge group have been reproduced. To address the case of bifundamental matter
coupled with multiple gauge groups, the construction of a new operator $V_{12}(\vec a_1, \vec a_2|m_{12}):\CV_{\vec a_2}\to\CV_{\vec a_1}$ is required. This operator intertwines two SH$^c$ representations specified by $\vec a_1, \vec a_2$, with a different rank $N_1$ for $\CV_{\vec a_1}$ and $N_2$ for $\CV_{\vec a_2}$,
\begin{equation}\label{intertwiner}
V_{12}(\vec a_1, \vec a_2|m_{12})=\sum_{\vec Y_1,\vec Y_2}\bZbf(\vec a_1,\vec Y_1;\vec a_2,\vec Y_2|m_{12})\ |\vec{a}_1, \vec{Y}_1\rangle\langle \vec a_2,\vec{Y}_2|,
\end{equation}
where a renormalized version of the bifundamental contribution has been used,
\begin{equation}\label{intertwiner1}
\bZbf(\vec a_1,\vec Y_1;\vec a_2,\vec Y_2|m_{12})=\sqrt{\Zv(\vec a_1,\vec Y_1)\Zv(\vec a_2,\vec Y_2)}\Zbf(\vec a_1,\vec Y_1;\vec a_2,\vec Y_2|m_{12}).
\end{equation}
Several algebraic properties of the intertwiner operator were studied from the viewpoint of SH$^c$ in \cite{Kanno2013}, in relation with a recursion formula satisfied by $\bZbf$.\footnote{In \cite{Kanno2013} it was assumed that the ranks of the two representations are the same. However, the computation performed there can be straightforwardly generalized to the case $N_1\neq N_2$.}

The intertwiner satisfies a set of identities which resemble
the conditions (\refOld{action_Gaiotto}--\refOld{action_Gaiotto_adj2}) for the Gaiotto states:

\begin{eqnarray}
&&D^{\vec a_1}_{-1}(z) V_{12}(\vec a_1, \vec a_2|m_{12})-
V_{12}(\vec a_1, \vec a_2|m_{12})D^{\vec a_2}_{-1}(z-m_{12})\nonumber\\
&&~~~~~~~~~~=\dfrac1{\sqrt{-\e_1\e_2}} \mathrm{P}^-_z\left(\dfrac1{\CY^{(1)}(z)}V_{12}(\vec a_1, \vec a_2|m_{12})\CY^{(2)}(z+\e_+-m_{12})\right),
\label{action_vertex1}\\
&&D^{\vec a_1}_{+1}(z) V_{12}(\vec a_1, \vec a_2|m_{12})-
V_{12}(\vec a_1, \vec a_2|m_{12})D^{\vec a_2}_{+1}(z+\e_+-m_{12})\nonumber\\
&&~~~~~~~~~~=-\dfrac1{\sqrt{-\e_1\e_2}} \mathrm{P}^-_z\left(\CY^{(1)}(z+\e_+)V_{12}(\vec a_1, \vec a_2|m_{12})\dfrac1{\CY^{(2)}(z+\e_+-m_{12})}\right).
\label{action_vertex2}
\end{eqnarray}
Here the notation $\CY^{(i)}$
($i=1,2$) represents the action of the vertex operor $\CY$ in the space $\CV_{\vec a_i}$. These formulas characterize the transformation of the bifundamental contribution under the action of SH$^c$. 
The proof of the formulae is summarized in
the appendix \refOld{AppA}.

In the 4D/2D correspondence, the intertwiner is described as a vertex operator of the form, $V=V^\mathrm{CO} V^\mathrm{Toda}$ where $V^\mathrm{CO}$ is the Carlsson-Okounkov vertex \cite{carlsson2008exts} for the $U(1)$ factor and $V^\mathrm{Toda}$ is the vertex operator of Toda field theory associated with the $W_N$ algebra. This construction, however, has some limitations.
One issue is the technical difficulty to define
the transformation properties of $V^\mathrm{Toda}$ for higher
spin generators. We have to face a nonlinear expression
in terms of $W$ generators or Toda fields which
is usually not manageable. A more serious issue is the impossibility to define an intertwiner between $W_N$ and $W_M$ 
Toda systems with $N\neq M$ since there is no obvious
correspondence between the generators in $W_N$ algebras with a different $N$ (see, for example \cite{Kanno:2010kj,Drukker:2010vg}, for attempts to explore such a setup).
At the level of SH$^c$, the correspondence between the generators
for representations of a different rank becomes obvious
and the transformation properties
\ref{action_vertex1} and \ref{action_vertex2}
are compact and tractable.
Furthermore, it was confirmed in \cite{Kanno2013, Matsuo2014}
that these conditions contain
the modified Ward identities for the U(1) current and 
the Virasoro operator for $V=V^\mathrm{CO} V^\mathrm{Toda}$
when $N_1=N_2$.
In this sense, our characterizations of the intertwiner 
is a natural generalization of the
conventional vertex operators in Toda field theories
to study the 4D/2D correspondence.

\paragraph{Gaiotto state from intertwiner} As briefly recalled in the previous subsection, the study of the instanton partition functions for miscellaneous field content can be reduced to the analysis of the bifundamental hypermultiplet. This fact has an important consequence for the SH$^c$ realization that we explain here. We first consider the special case $N_1=N$, $N_2=0$ and $m_{12}=0$ for the intertwiner. Here the rank $0$ representation means a trivial representation which consists of one state -- the vacuum $|,\rangle$ (empty slots means that we have no Fock space). 
Since $\Zbf(\vec a,\vec Y;,|0)=\Zv(,)=1$, we find after omitting the trivial bra vacuum,
\begin{eqnarray}
V_{12}(\vec a, |0)=|G,\vec{a}\rangle.
%\quad V_{12}(,\vec a|0)=\langle G, \vec a|\,.
\end{eqnarray}
Taking the opposite case of a rank zero representation for the first node produces the bra Gaiotto state in a similar way. 
In this sense, the recursion properties of the Gaiotto states \ref{action_Gaiotto} -- \ref{action_Gaiotto_adj2} are straightforward consequences of \ref{action_vertex1} and \ref{action_vertex2}. We note that we can take $D^{(2)}_\eta(z)=0$ for the trivial representation and the action operator $\CY$ on $\CV_2$ is replaced by $1$.

Inclusion of (anti-)fundamental hypermultiplet is also straightforward.  From \ref{def_Zf} and \ref{def_Zaf}, after fixing $N_1=N$, $N_2=\tilde{N}$, $m_{12}=0$ and $\vec{a}_2=\vec{m}+\e_+$, we obtain that the action of the intertwiner on the vacuum produces the Gaiotto state with a 
flavor vertex operator inserted:
\begin{eqnarray}\label{I2G}
V_{12}(\vec{a},\vec{m}+\e_+|0)|\vec{m}+\e_+,\vec{\emptyset}\rangle
=U(\vec{m})|G,\vec{a}\rangle\,.
\end{eqnarray}

\begin{figure}[bpt]
\begin{center}
\begin{tikzpicture}[scale=0.8]
\filldraw [fill=yellow!30] (0,0) circle (0.9 cm);
\draw [->] (-1,0.3) .. controls (-3,3) and (3,3) .. (1,0.3);
\draw (0,2.8) node {bifund. $m$};
\draw (0,0) node {$SU(N)$};
\draw (1.2,-0.8) node {vevs $\vec{a}$};
%\draw (2,0) node {adj. $m$};
\end{tikzpicture}	
\end{center}
\caption{$A^{(0)}_1$ quiver}
\label{f:A01}
\end{figure}
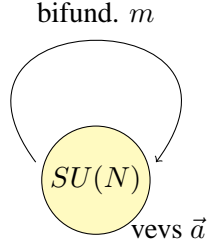

\paragraph{Full partition function} We now have all the elements to write down the instanton partition function of any linear quiver as a product of operators,
{\small
\begin{eqnarray}
\CZ_\text{inst}&=& \left\{\begin{array}{ll}
\langle G,\vec a_1| q_1^D U(\vec m_1)V_{12}(\vec a_1, \vec a_2|m_{12})q_2^D U(\vec m_2)V_{23}(\vec a_2, \vec a_3|m_{23}) \cdots|G,\vec a_{\nQ}\rangle & \mbox{for } A_{\nQ}\,,\\
\mbox{Tr}_{\CV_{\vec a_1}} \left[q_1^D U(\vec m_1)V_{12}(\vec a_1, \vec a_2|m_{12})q_2^D U(\vec m_2)V_{23}(\vec a_2, \vec a_3|m_{23}) \cdots  V_{\nQ 1}(\vec a_{\nQ},\vec a_1|m_{\nQ 1}) \right] &\mbox{for } A^{(1)}_{\nQ}.
\end{array} 
\right.\label{Nek_cor}
\end{eqnarray}
}
To each arrow $i\to j$ of the quiver is associated the intertwiner $V_{ij}(\vec a_i,\vec a_j|m_{ij})$, and to each node $i$ an operator $q_i^D U(\vec m_i)$. For the linear quivers, the resulting operator is sandwiched between the Gaiotto states attached to the first and the last node. On the contrary, a trace is directly obtained for the affine quiver from the intertwiners, it is defined as
\begin{equation}
\mbox{Tr}_{\CV_{\vec a}}\cdots=\sum_{\vec Y}\langle\vec a,\vec Y|\cdots\aY.
\end{equation}
As an example, the partition function of $\mathcal{N}=2^\ast$ theory represented in figure \refOld{f:A01} with bifundamental fields of mass $m$ reads
\begin{equation}\label{A:1:1}
\CZ_\text{inst}=\mbox{Tr}_{\CV_{\vec a}} \left[q^D V_{11}(\vec a, \vec a|m)\right]=\sum_{\vec Y}q^{|\vec Y|}\Zv(\vec a, \vec Y)\Zbf(\vec a,\vec Y;\vec a,\vec Y|m)=\sum_{\vec Y}q^{|\vec Y|}\Zv(\vec a, \vec Y)
\CZ_\mathrm{adj}(\vec a,\vec Y|m),
\end{equation}
thanks to the orthonormality property of the states $\aY$.

\paragraph{Alternative expressions} Although it will not be used in this paper, we would like to provide, as a side remark, a new expression for the bifundamental contribution \ref{Zbfd} involving the vertex operator $\CY(z)$. This expression is a consequence of the property \ref{prop_Zbf1} expressing the variation of $\Zbf(\vec a,\vec Y;\vec b,\vec W|m)$ under the addition of a box in the Young diagrams $\vec Y$. It turns out that the right hand side of \ref{prop_Zbf1} is independent of the actual content of boxes in the Young diagrams $\vec Y$. As a result, this formula can be used to build recursively $\Zbf(\vec a,\vec Y;\vec b,\vec W|m)$ from $\Zbf(\vec a,\vec \emptyset;\vec b,\vec W|m)$, adding boxes one by one. Since, $\Zbf(\vec a,\vec \emptyset;\vec b,\vec W|m)$ can be further identified with a fundamental contribution of mass $\vec a-m$, it is shown that
\begin{align}
\begin{split}
 \Zbf(\vec a,\vec Y;\vec b,\vec W|m)&=\langle \vec b,\vec W|U(\vec a-m)| \vec b,\vec W\rangle\prod_{x\in \vec Y}\la \vec b,\vec W|\CY(\phi_x-m+\e_+)|\vec b,\vec W\ra\\
&=\langle \vec a,\vec Y|U(\vec b+m-\e_+)| \vec a,\vec Y\rangle\prod_{x\in \vec W}\la \vec a,\vec Y|\CY(\phi_x+m)|\vec a,\vec Y\ra
\end{split}
\end{align}
where the second equality has been obtained by exploiting the symmetry under the exchange of $(\vec a,\vec Y)\leftrightarrow(\vec b,\vec W)$ and $m\leftrightarrow\e_+-m$.  As a special case of this expression, new formulae for the vector contribution and the Gaiotto states can also be deduced,
\begin{equation}\label{Zv1}
\Zv(\vec a,\vec Y)=\la\vec a,\vec Y\Big|U(\vec a-\e_+)^{-1}\prod_{x\in\vec Y}\CY(\phi_x)^{-1}\Big|\vec a,\vec Y\ra,\quad|G,\vec a\rangle=\sum_{\vec Y}\dfrac1{\sqrt{U(\vec a-\e_+)}}\prod_{x\in Y}\dfrac1{\sqrt{\CY(\phi_x)}}\aY.
\end{equation}

% \begin{align}
% \begin{split}
% &\Zv(\vec a,\vec Y)=\la\vec a,\vec Y\Big|U(\vec a-\e_+)^{-1}\prod_{x\in\vec Y}\CY(\phi_x)^{-1}\Big|\vec a,\vec Y\ra,\\
% &|G,\vec a\rangle=\sum_{\vec Y}\dfrac1{\sqrt{U(\vec a-\e_+)}}\prod_{x\in Y}\dfrac1{\sqrt{\CY(\phi_x)}}\aY.
% \end{split}
% \end{align}

\section{Ward identities of SH$^c$ and qq-character}
In the previous section, we have seen that the instanton
partition function for any $A_Q$ type quiver gauge theories
can be written by combining Gaiotto states,
dilatation operators,
flavor vertex operators and intertwiners as in \ref{Nek_cor}.
The behavior of these states/operators under the action of SH$^c$ generators has been characterized through the set of relations (\refOld{action_Gaiotto}--\refOld{action_Gaiotto_adj2}), 
\ref{dilatation}, \ref{UDU3} and
(\refOld{action_vertex1}--\refOld{action_vertex2}). As a side result, one obtains
a series of consistency conditions by inserting $D_{\pm 1}(z)$
in the correlator and evaluating the inner product in
two different ways,
\begin{eqnarray}\label{WI}
\left(\langle G,\vec{a}_1|\mathcal{O}_1 D_{\pm 1}(z)\right)\mathcal{O}_2| G,\vec{a}_Q\rangle
=\langle G,\vec{a}_1|\mathcal{O}_1 \left(D_{\pm 1}(z)\mathcal{O}_2 |G,\vec{a}_Q\rangle
\right)\,,
\end{eqnarray}
where $\mathcal{O}_i$ denotes a combination of flavor vertex operator,
dilatation operators and intertwiners.
These conditions may be regarded as the Ward identities
for the correlation functions of SH$^c$.  Since it is written
as a generating function with parameter $z$, it gives an infinite number
of constraints.
In the following, we evaluate the explicit
form of these identities.  We observe that their structure takes the form of a double quantum deformation of the character
formulae for $A_Q$, the so-called \textit{qq-character} proposed by
Nekrasov, Pestun and Shatashvili \cite{Nekrasov:2012xe, Nekrasov2015}. In the next section, we will discuss another interpretation of these formulae as a quantum deformation of the Seiberg-Witten curve.

\subsection{$A_1$ quiver}
%A direct application of our formalism is the derivation of the qq-characters. This double deformation of the $sl(2)$ characters were  For simplicity, we first restrict ourselves to the case of a single node. The corresponding gauge theory has an $SU(N)$ gauge group and $\tN$ fundamental flavors with masses $m^{(f)}$.The qq-characters are operators $\chi_r(z):\CV_{\vec a}\to \CV_{\vec a}$ that evaluates to a polynomial in the expectation value of a (normalized) Gaiotto state,

We start from the expression \ref{Zinst_A1_pure} of the instanton partition function for pure SYM with $SU(N)$ gauge group, and consider
the insertion of the operator of $D_{-1}(z)$,
$
%\langle G,\vec a|D_{-1}(z)q^{D}U(\vec m)|G,\vec a\rangle
\langle G,\vec a|D_{-1}(z)q^{D}|G,\vec a\rangle,
$
evaluated in two different ways as in \ref{WI}. After the use of the identities
(\refOld{dilatation}, \refOld{action_Gaiotto}, \refOld{action_Gaiotto_adj}),
we arrive at,
\begin{eqnarray}\label{qg_A1}
%\langle G,\vec{a}|\mathrm{P}_z^-\left(\CY(z+\e_+)+qm(z)\frac{1}{\CY(z)}\right)q^D 
%U(\vec m)|G,\vec a\rangle=0\,,\quad m(z)=\prod_{f=1}^{\tN}(z-m^{(f)}),
\langle G,\vec{a}|\mathrm{P}_z^-\left(\CY(z+\e_+)+\frac{q}{\CY(z)}\right)q^D 
|G,\vec a\rangle=0\,.
\end{eqnarray}
%where $m(z)$ denotes the mass polynomial.
%
The insertion of $\CY$ has the effect of adding extra
factors to the partition function.  For example,
from \ref{prop_CY},
\begin{eqnarray}
\langle G,\vec{a}|\CY(z+\e_+)q^D |G,\vec a\rangle
=\sum_{\vec Y}q^{|\vec Y|}\left(\dfrac{\prod_{x\in A(\vec{Y})}(z+\e_+-\phi_x)}{\prod_{x\in R(\vec{Y})}(z-\phi_x)}
\right)
\Zv(\vec a, \vec Y).
%\langle G,\vec{a}|\CY(z+\e_+)q^D U(\vec m)|G,\vec a\rangle
%=\sum_{\vec Y}q^{|\vec Y|}\left(\dfrac{\prod_{x\in A(\vec{Y})}(z+\e_+-\phi_x)}{\prod_{x\in R(\vec{Y})}(z-\phi_x)}
%\right)
%\Zv(\vec a, \vec Y)\Zf(\vec m;\vec a,\vec Y).
\end{eqnarray}
We will use the following notation for the expectation value
of the Gaiotto state:
\begin{equation}\label{def_avr}
\la\cdots\ra=\dfrac1{\Zi}\sum_{\vec Y}q^{|\vec Y|}\Zv(\vec a,\vec Y)\langle\vec a,\vec Y|\cdots\aY.
%\la\cdots\ra=\dfrac1{\Zi}\sum_{\vec Y}\Zv(\vec a,\vec Y)\Zf(\vec m;\vec a,\vec Y)\langle\vec a,\vec Y|\cdots\aY.
\end{equation}
This defines an average of operators acting on states $\aY$ that is normalized to $\la1\ra=1$. 
The relation \ref{qg_A1} is rewritten in the form:
\begin{equation}\label{PYq/Y}
\mathrm{P}_z^-\la \CY(z+\e_+)+\frac{q}{\CY(z)}\ra =0\,.
\end{equation}
This condition is the generating function
of an infinite number of constraints
on the instanton partition function.
At the same time, this formula implies that
\begin{equation}\label{def_chi1}
\chi(z):=\la\CY(z+\e_+)\ra+\la\dfrac{q}{\CY(z)}\ra
=\la \mathrm{P}^+_z(\CY(z+\e_+))\ra
\end{equation}
has no negative powers of $z$ in the Laurent expansion
at $z=\infty$.  We note that $\CY(z)$ behaves
as $\CY(z)\sim z^N$
as $z\rightarrow \infty$.  It implies that $\chi(z)$
thus defined is a polynomial in $z$ of degree $N$. The expression $\chi\sim y+1/y$ is reminiscent of the character of $sl(2)$ for the fundamental representation.
The formula \ref{def_chi1} is deformed by two
parameters $\e_{1,2}$ and was referred as a fundamental
qq-character in \cite{Nekrasov:2012xe, Nekrasov2013, Nekrasov2015}
for the quantum deformed Yangian $Y_{\e}(sl(2))$.

The inclusion of 
fundamental hypermultiplets with $\tilde{N}$
flavor is a straightforward generalization. The only necessary modification is to insert a flavor vertex operator
$U(\vec m)$ in front of $|G,\vec{a}\rangle$.
The commutator with $D_{-1}(z)$ is obtained from \ref{com_D_Phi}:
\begin{equation}\label{DU}
D_{-1}(z)U(\vec m)=U(\vec m)\mathrm{P}^-_z\left[D_{-1}(z)m(z)\right],
\quad m(z)=\prod_{f=1}^{\tilde{N}}(z-m^{(f)})\,.
\end{equation}
%
%The fundamental qq-character is defined as
%The SH$^c$ algebra offers a way to derive an alternative expression for the vev of the fundamental qq-character that allows us to show its polynomiality property. For simplicity, we first consider the case of pure $\mathcal{N}=2$ SYM. The alternative expression is obtained by considering the operator $D_{-1}(z)$ between two Gaiotto states. The commutation relation between $D_{-1}(z)$ and $q^D$ implies
%\begin{equation}
%\bGa D_{-1}(z)q^D\Ga=\bGa q^{D+1}D_{-1}(z)\Ga.
%\end{equation}
%In the RHS, the correlator is evaluated using the action \ref{action_Gaiotto} of the operator $D_{-1}(z)$ on the Gaiotto state, whereas the LHS is obtained from the left action \ref{action_Gaiotto_adj} on the bra, leading to the identity
%\begin{equation}
%q\la\dfrac1{\CY(z)}\ra=-\mathrm{P}^-_z\la \CY(z+\e_+)\ra.
%\end{equation}
%Let $\chi(z)=\mathrm{P}^+_z\CY(z+\e_+)$ be the sum of the positive powers in the expansion of the operator $\CY(z+\e_+)$ at infinity. The previous identity implies
%\begin{equation}
%\la\chi(z)\ra=\la\CY(z+\e_+)\ra-\mathrm{P}^-_z\la\CY(z+\e_+)\ra=\la\CY(z+\e_+)\ra+q\la\dfrac1{\CY(z)}\ra.
%\end{equation}
%Thus we have shown that $\chi(z)$ and $\chi_1(z)$ have the same average, and since by definition $\chi(z)$ is a polynomial in $z$, so is $\la\chi_1(z)\ra$.
%
Inserting this relation between two Gaiotto states (with an operator $q^D$), and then evaluating the action of $D_{-1}(z)$ through \ref{action_Gaiotto} and \ref{action_Gaiotto_adj} leads to
\begin{equation}\label{chi_mass_pm}
\mathrm{P}^-_z\la \CY(z+\e_+)+q\dfrac{m(z)}{\CY(z)}\ra=0,
\end{equation}
where the average acquired an extra factor $\Zf(\vec m;\vec a,\vec Y)$,
\begin{equation}\label{avr_mass}
 \la\cdots\ra=\dfrac1{\Zi}\sum_{\vec Y}q^{|\vec Y|}\Zf(\vec m;\vec a,\vec Y)\Zv(\vec a,\vec Y)\langle\vec a,\vec Y|\cdots\aY.
\end{equation}
After including the fundamental hypermultiplets, the qq-character is modified to
\begin{equation}\label{equ_qq}
\chi(z)=\la\CY(z+\e_+)+\dfrac{qm(z)}{\CY(z)}\ra\,.
%=\mathrm{P}^+_z\la\CY(z+\e_+)+\dfrac{qm(z)}{\CY(z)}\ra\,.
\end{equation}
As a consequence of \ref{chi_mass_pm}, $\mathrm{P}^-_z\chi(z)=0$ and the qq-character is again a polynomial of degree $N$ in $z$. A more detailed discussion of the qq-character in the presence of fundamental hypermultiplets is presented in appendix \refOld{AppB}.

%There, it is shown that the character $\chi(z)$ equals the average of the operator $\mathrm{P}^+_z\CY(z+\e_+)$, which is a direct consequence of \ref{chi_mass_pm} if $\tilde{N}<N$, but also holds in the general case.

% As long as the gauge theory is asymptotic free or conformal
% ($N_f\leq 2N$), the qq-character $\chi(z)$ is a polynomial
% of order $N$.
% A more detailed treatment of the system with the fundamental
% multiplet $\tN\leq 2N$ can be found in the appendix \refOld{AppB}. 
%It turns out that for any number of fundamental flavors, the vev of the fundamental character is always equal to $\la\chi(z)\ra$. In addition, we would like to stress that the same expressions would have been obtained using the generators of positive degree $D_{1}(z)$ instead of the negative ones.

In the case $\tilde{N}<N$, the ratio $m(z)/\CY(z)$ has no polynomial part and the character $\chi(z)$ equals the average of the operator $\mathrm{P}^+_z\CY(z+\e_+)$. As a result, explicit expressions for the qq-character can be obtained by expansion of $\CY(z+\e_+)$ at infinity using the properties (\refOld{def_CY},\refOld{prop_CY}):
\begin{eqnarray}
\CY(z+\e_+)&=&\prod_{\ell=1}^{N}(z+\e_+-a_\ell)\left(
1-\e_1\e_2\frac{d}{dz} D_0(z)+\mbox{higher terms in } \e
\right)\nonumber\\
\label{exp_CY}
&=& \prod_{\ell=1}^{N}(z+\e_+-a_\ell)
%z^N +z^{N-1}\sum_{\ell=1}^N (\e_+-a_\ell)
+\e_1\e_2 z^{N-2}\ D+O(z^{N-3}).
\end{eqnarray}
In the average \ref{def_avr} the operator $D$ with eigenvalue $|\vec Y|$ can be replaced by a logarithmic $q$-derivative,
\begin{equation}
\chi(z)=\prod_{\ell=1}^{N}(z+\e_+-a_\ell)+\e_1\e_2z^{N-2}q\p_q\log\mathcal{Z}_\text{inst}+O(z^{N-3})\,.
\end{equation}
Specializing to $N=1$ and to $N=2$ with $a_1=-a_2=a$, we deduce the following expressions
\begin{align}
\begin{split}\label{chi1_expl}
&U(1):\quad\ \chi(z)=z+\e_+-a,\\
&SU(2):\quad \chi(z)=(z+\e_+)^2-a^2+\e_1\e_2q\p_q\log\mathcal{Z}_\text{inst}.
\end{split}
\end{align}

\subsection{qq-characters of higher representations for the $A_1$ quiver}
In a series of recent lectures, Nekrasov
proposed a generalization of the qq-character for higher representations of $Y_{\e}(sl(2))$ \cite{Nekrasov2015}.
Higher qq-characters involve a set of complex parameters $\nu_1,\cdots,\nu_r\in\mathbb{C}$, and they are defined as
\begin{equation}\label{higher-qq}
\chi_r(z|\nu_1,\cdots \nu_r)=\sum_{I\sqcup J=\{1,\cdots,r\}}q^{|J|}
\prod_{\superp{i\in I}{j\in J}} S(\nu_i-\nu_j)
\la\prod_{i\in I}\CY(z+\e_++\nu_i)\prod_{j\in J}\dfrac{m(z+\nu_j)}{\CY(z+\nu_j)}\ra .
\end{equation}
Here $S(z)$ is the scattering factor \ref{braiding}.
It is claimed in \cite{Nekrasov2015} that the expectation value of these operators is again a polynomial in $z$. This proposal has been verified using our formalism in appendix \refOld{AppB} for the second character of pure $SU(N)$ SYM in the restricted cases $N=1$ and $N=2$. The second character can be rewritten using the shifted spectral variables $z_1=z+\nu_1$, $z_2=z+\nu_2$,
\begin{align}
\begin{split}\label{chi2}
\chi_2(z_1,z_2)&=\la\left(\CY(z_1+\e_+)+\dfrac{qm(z_1)}{\CY(z_1)}\right)
\left(\CY(z_2+\e_+)+\dfrac{qm(z_2)}{\CY(z_2)}\right)\ra\\
&+q\dfrac{\e_1\e_2}{z_{12}}\la\dfrac{\CY(z_1+\e_+)}{z_{12}+\e_+}\dfrac{m(z_2)}{\CY(z_2)}+\dfrac{\CY(z_2+\e_+)}{z_{12}-\e_+}\dfrac{m(z_1)}{\CY(z_1)}\ra,
\end{split}
\end{align}
with $z_{12}=z_1-z_2$. This condition is equivalent to \ref{higher-qq}.
From the insertion of two operators $D_{1}(z_1)D_{-1}(z_2)$ within two Gaiotto states, it is possible to show that
\begin{equation}
\mathrm{P}_{z_1}^-\mathrm{P}_{z_2}^-\chi_2(z_1,z_2)=0.
\end{equation}
The polynomiality is further obtained in the cases $N=1$ and $N=2$ by employing the explicit expression \ref{exp_CY} of the operator $\CY(z+\e_+)$. In both cases, it was found that
\begin{equation}
\chi_2(z_1,z_2)=\mathrm{P}_{z_1}^+\mathrm{P}_{z_2}^+\la\CY(z_1+\e_+)\CY(z_2+\e_+)\ra+\dfrac{2q\e_1\e_2}{z_{12}^2-\e_+^2}.
\end{equation}

\subsection{Generalization to the $A_Q$-type quiver}
For simplicity here we will only treat explicitly the case of the $A_2$ quiver without fundamental matter fields. For any operator $\Op$ we introduce the index $\a=1,2$ labeling the space $\CV_{\vec a_\a}$ in which the operator acts, and we associate the expectation value
\begin{equation}\label{A2_average}
\la\Op^{(\a)}(z)\ra=\dfrac1{\Zi}\sum_{\vec Y_1,\vec Y_2}q_1^{|\vec Y_1|}q_2^{|\vec Y_2|}\Zv(\vec a_1,\vec Y_1)\Zv(\vec a_2,\vec Y_2)\Zbf(\vec a_1,\vec Y_1;\vec a_2,\vec Y_2|m_{12})\ \langle \vec a_\a,\vec Y_\a|\Op_\a(z)|\vec a_\a,\vec Y_\a\rangle.
\end{equation} %In the quiver case, a fundamental character $\chi_1^{(i)}(z)$ can be associated to each node $i$ following \ref{def_chi1}. We also define for each node $i$ a polynomial operator $\chi^{(i)}(z)=\mathrm{P}^+_z\CY_i(z+\e_+)$ acting in the representation space $\CV_{\vec a_i}$. 
To derive the qq-character relations, we consider the commutation relation \ref{action_vertex1} between the SH$^c$ generating series $D_{\eta}(z)$ and the intertwiner operator. 
%Inserting this relation between two Gaitto states associated to each node, further decorated by the instanton counting operator $q^{D}$ as in
We consider the operator insertion of the following type,
\begin{equation}
\langle G,\vec a_1|D^{\vec a_1}_{-1}(z)q_1^{D}V_{12}(\vec a_1, \vec a_2|m_{12})q_2^{D}|G,\vec a_2\rangle,\quad \langle G,\vec a_1|q_1^{D}V_{12}(\vec a_1, \vec a_2|m_{12})q_2^{D}D^{\vec a_2}_{+1}(z)|G,\vec a_2\rangle,
\end{equation}
and then using the action of the SH$^c$ modes on Gaiotto states, it is possible to derive the following identity, obtained respectively from the former and latter expressions:%\footnote{Note that in the case $N_1=N_2$, we can simplify
%\begin{equation}
%\mathrm{P}_z^-\la\dfrac{\CY^{(2)}(z+\e_+-m_{12})}{\CY^{(1)}(z)}\ra=\la\dfrac{\CY^{(2)}(z+\e_+-m_{12})}{\CY^{(1)}(z)}\ra-1.
%\end{equation}}
\begin{align}
\begin{split}
&\mathrm{P}_z^-\la
\CY^{(1)}(z+\e_+)+q_1\dfrac{\CY^{(2)}(z+\e_+-m_{12})}{\CY^{(1)}(z)}
+q_1q_2\dfrac1{\CY^{(2)}(z-m_{12})}\ra=0,\\
&\mathrm{P}_z^-\la
\CY^{(2)}(z+\e_+)+q_2\dfrac{\CY^{(1)}(z+m_{12})}{\CY^{(2)}(z)}
+q_1q_2\dfrac1{\CY^{(1)}(z+m_{12}-\e_+)}\ra=0.
\end{split}
\end{align}
These identities imply that the two following qq-characters are polynomials in $z$:
\begin{align}
\begin{split}\label{A2_qq_bis}
\chi^{(1)}(z)&=\la
\CY^{(1)}(z+\e_+)+q_1\dfrac{\CY^{(2)}(z+\e_+-m_{12})}{\CY^{(1)}(z)}
+q_1q_2\dfrac1{\CY^{(2)}(z-m_{12})}\ra,\\
\chi^{(2)}(z)&=\la
\CY^{(2)}(z+\e_+)+q_2\dfrac{\CY^{(1)}(z+m_{12})}{\CY^{(2)}(z)}
+q_1q_2\dfrac1{\CY^{(1)}(z+m_{12}-\e_+)}\ra.\\
\end{split}
\end{align}
Generalization of these formulae to the $A_Q$ quiver with fundamental
multiplets is straightforward.
For example, the first one is generalized to 
\begin{equation}
\chi^{(1)}(z)=\sum_{i=1}^{Q+1}\left[\prod_{j=1}^{i-1} q_jm_j(z-\z_j)\right]\la\dfrac{\CY_i(z+\e_+-\z_i)}{\CY_{i-1}(z-\z_{i-1})}\ra,\quad \z_j=\sum_{k=1}^{j-1}m_{k,k+1},
\end{equation}
% \begin{eqnarray}
% \chi^{(1)}(z)&=
% \sum_{i=1}^{Q+1} (\prod_{j=1}^{i-1} q_jm_j(z))\la 
% \dfrac{\CY_i(z+\e_+-\sum_{j=1}^i m_{j,j+1})}{\CY_{i-1}(z-\sum_{j=1}^{i-1} m_{j,j+1})}\ra,
% \end{eqnarray}
with $\z_0=\z_1=0$, $\CY_0=\CY_{Q+1}:=1$ and the mass polynomials $m_i(z)=\prod_{f=1}^{\tilde N_i}(z-m^{(i)}_f)$ associated to the fundamental multiplet of the node $i$, with flavor group $SU(\tilde{N}_i)$.
We note that for a linear quiver with $N=N_1\geq N_2\geq \cdots \geq N_Q$,
the rank of the flavor group need to satisfy $\tilde N_i\leq 2 N_i-N_{i+1}-N_{i-1}$
with $N_0=N_{Q+1}=0$.  In this set-up, the character $\chi^{(1)}(z)$ is a polynomial of degree at most $N$.

As already stated, in the weak coupling limit $q_2\to0$ the second node of the quiver diagram acts as a set of $N_2$ fundamental flavors of mass $a_\ell^{(2)}$ coupled to the first node. This relation can also be observed at the level of characters. As can be seen from \ref{A2_average} in this limit only the empty $N_2$-tuple $\vec Y_2=\vec \emptyset$ contribute to the sum, $\Zv(\vec a_2,\vec\emptyset)=1$ and $\Zbf\to\Zf$. We further notice that the operator $\CY^{(2)}(z+\e_+)$ becomes polynomial and, as such, can be identify with $\chi^{(2)}(z)$, it reproduces a mass polynomial with masses $a_\ell^{(2)}-\e_+$,
\begin{equation}
\chi^{(2)}(z)\to \prod_{\ell=1}^{N_2}(z-a_\ell^{(2)}+\e_+)=:m(z).
\end{equation}
In this weak coupling limit, the first equation in \ref{A2_qq_bis} becomes the equation \ref{equ_qq} for the massive qq-character, with an extra shift of the fundamental masses by $m_{12}$.

%\subsection{Analog for $A^{(1)}_Q$ quiver}
%For the quiver gauge theories of affine type, the analog of the Ward identity
%does not close in the finite terms.  In order to see it, let us consider first the
%simplest $A^{(1)}_1$ case.  As in the $A_Q$ quiver, we insert $D_{\pm 1}(z)$
%in the correlator.  We consider the following insertion,
%$
%\mbox{Tr}_{\mathcal{V}_a}\left(
%D_{-1}(z) q^D V(\vec a,\vec a|m)
%\right)
%$.
%We move $D_{-1}$ to the right using \ref{dilatation},\ref{action_vertex1} to obtain
%\begin{eqnarray}
%&& \mbox{Tr}_{\mathcal{V}_a}\left(
%D_{-1}(z) q^D V(\vec a,\vec a|m)\right)\nonumber\\
%&&~~~~~= \mbox{Tr}_{\mathcal{V}_a}
%\left(
%q^{D+1} \mathrm{P}^+_z\left(\frac{1}{\CY(z)}V(\vec a,\vec a|m)\CY(z+\e_+-m)\right)
%+ q^{D+1} V(\vec a,\vec a|m)D_{-1}(z-m)
%\right)\,.
%\end{eqnarray}
%The $D_{-1}(z-m)$ operator in the second term
%can be moved in front of $q^{D+1} V(\vec a,\vec a|m)$ by using the cyclicity
%in the trace.  One may continue this operation and obtain:
% \begin{eqnarray}
% \mbox{Tr}_{\mathcal{V}_a}\left(
% D_{-1}(z) q^D V(\vec a,\vec a|m)\right)
% &=& \sum_{\ell=1}^\infty q^\ell \mathrm{P}^+_z
% \la
% \frac{\CY(z+\e_+-\ell m)}{\CY(z-(\ell-1)m)}
% \ra\,.
% \end{eqnarray}
%Similarly we rotate $D_{-1}(z)$ to left and we obtain...

\section{Quantum Seiberg-Witten geometry}
In the limit $\e_1,\e_2\to0$, the Omega-background reduces to $\mathbb{R}^4$ and the infrared theory is characterized by a complex algebraic curve. This curve, together with a differential form, determines the prepotential of the theory through the Seiberg-Witten relations. It is also associated to the spectral curve of a classical integrable system in the Bethe/gauge correspondence (see for instance \cite{Marshakov1999} and references inside). For simplicity here, we focus our discussion on the case of a single node with gauge group $SU(N)$ and a number $\tilde{N}$ of fundamental multiplets. In this case, the algebraic curve can be written in the form
\begin{equation}\label{SW_curve}
y+q\dfrac{m(z)}{y}=\prod_{\ell=1}^N(z-a_\ell).
\end{equation}
This expression should be compared with the definition \ref{def_chi1} of the qq-character. It is then appealing to interpret the qq-character as a double deformation of the Seiberg-Witten geometry, where the expectation value of the operator $\CY(z)$ reduces to the complex parameter $y$ of the curve $E(y,z)=0$, while the qq-character $\chi(z)$ reproduces the gauge polynomial in the RHS of \ref{SW_curve}. This is indeed the case, as we will demonstrate shortly.

The discussion becomes even more illuminating if we introduce the intermediate background $\mathbb{R}^2_{\e_1}\times\mathbb{R}^2$ obtained in the Nekrasov-Shatashvili limit $\e_2\to0$ of the Omega-background. This $\e_1$-deformation of the Euclidean background is known to be responsible for the quantization of the classical integrable system associated to the $\mathcal{N}=2$ gauge theory \cite{Nekrasov2003a}. In this background, the Seiberg-Witten curve is replaced by a Baxter TQ-equation that has been derived in \cite{Poghossian2010,Fucito2011} (the derivation was later extended to quivers in \cite{Fucito2012,Nekrasov2013}),
\begin{equation}\label{def_q}
T(z)Q(z)=Q(z+\e_1)+qm(z)Q(z-\e_1),\quad Q(z)=\prod_r(z-u_r),
\end{equation}
where $T(z)$ and $Q(z)$ denote respectively the Baxter T- and Q-polynomials. The TQ-equation can be recast in a form more similar to the original Seiberg-Witten curve \ref{SW_curve} by the introduction of the ratio $Y(z)=Q(z)/Q(z-\e_1)$:\footnote{The TQ-equation can also be written in an operatorial form,
\begin{equation}
\left(\hat y+qm(z)\hat y^{-1}\right)Q(z)=T(z) Q(z),\quad \hat y=e^{\e_1\p_z}
\end{equation}
where $\hat y$ is a shift operator. Here the non-commutativity of the variables $\hat y$ and $z$ becomes manifest and the previous relation defines a quantum curve. This difference equation is actually equivalent to a Schr\"odinger equation under a quantum change of variables \cite{Bourgine2012a}. This correspondence goes under the name of \textit{bispectral duality} \cite{Mironov2009b,Mironov2010a,Mironov2010} and can be seen as a degenerate version of the AGT correspondence relating the gauge theory in the NS background with the semiclassical Liouville/Toda theory.}
\begin{equation}\label{tq_rel}
T(z)=Y(z+\e_1)+q\dfrac{m(z)}{Y(z)}.
\end{equation}
In this form, it readily reproduces \ref{SW_curve} in the limit $\e_1\to0$. In order to show that the qq-character defines a sort of second quantization of the Seiberg-Witten geometry, we will take the NS limit and reproduce the TQ-relation \ref{tq_rel}, the operator $\CY(z)$ being reduced to the rational function $Y(z)$, and the qq-character to the T-polynomial.

To perform the NS limit, we will follow the procedure described in \cite{Bourgine2014a} (see also \cite{Nekrasov2013}) and first re-derive the Bethe equations. In the NS limit, the sum over Young diagrams entering the expression \ref{Zinst_A1} of the partition function is dominated by a Young diagram $\vec Y^\ast$ with infinitely many boxes.\footnote{This argument is similar to the one employed by Nekrasov and Okounkov in \cite{Nekrasov2003a} to perform the limit $\e_1,\e_2\to0$. The main difference here is that the critical Young diagram doesn't have a continuous profile but is instead described by a step-function where the plateaux are given by the Bethe roots.} This critical Young diagram minimizes the summation and its profile is obtained by solving the discrete saddle point equations:
\begin{equation}
\dfrac{q^{|\vec Y^\ast+x|}\Zv(\vec a,\vec Y^\ast+x)\Zf(\vec m;\vec a,\vec Y^\ast+x)}{q^{|\vec Y^\ast|}\Zv(\vec a,\vec Y^\ast)\Zf(\vec m;\vec a,\vec Y^\ast)}=1,\quad\forall x\in A(\vec Y^\ast).
\end{equation}
Taking into account the variation of the vector and fundamental contributions, we find
\begin{equation}\label{pre_Bethe}
-\dfrac{q}{\e_1\e_2}m(\phi_x)\dfrac{\prod_{y\in R(\vec Y^\ast)}(\phi_x-\phi_y)(\phi_x-\phi_y-\e_+)}{\prod_{\superp{y\in A(\vec Y^\ast)}{y\neq x}}(\phi_x-\phi_y)(\phi_x-\phi_y+\e_+)}=1,\quad\forall x\in A(\vec Y^\ast).
\end{equation}
Following \cite{Bourgine2014a}, we now consider only Young diagrams with infinitely high columns, and such that a box can be added to (or removed from) each column. Up to $\e_2$-corrections, the images under $\phi_x$ of a box $x\in A(\vec Y^\ast)$ and the box immediately below $x'\in R(\vec Y^\ast)$ are equal, they define the set of Bethe roots $u_r=\phi_x$ for $x\in R(\vec Y^\ast)$.\footnote{In a Young diagram $\l$, the image $\phi_x$ of $x=(i,\l_i)\in R(\l)$ is given explicitly by $\phi_x=a+(i-1)\e_1+(\l_i-1)\e_2$, it is finite in the limit $\e_2\to0$ since $\l_i$ tends to infinity such that $\e_2\l_i$ remains finite.} This is true for all boxes $x\in A(\vec Y^\ast)$, except for $N$ extra boxes (one for each diagram) that lie on the top right of the diagrams, and for which $\phi_x=\xi_\ell$ with $\xi_l=a_\ell+n_\ell\e_1$ and $n_\ell$ the number of columns for the Young diagram $Y_\ell^\ast$.It is emphasized that these extra boxes are necessary to fulfill the relation $|A(\vec Y^\ast)|=|R(\vec Y^\ast)|+N$ between the cardinal of the two sets. The number of columns $n_\ell$ in each diagram will play the role of a cut-off sent to infinity at the end of the computation. Under this identification, and taking into account the factor $-\e_1\e_2$ from the box $y$ of coordinate $\phi_y=\phi_x-\e_2$ just below $x$, we find in the limit $\e_2\to0$:
\begin{equation}\label{Bethe}
1=q\dfrac{m(u_r)}{\Xi(u_r)\Xi(u_r+\e_1)}\prod_{\superp{s=1}{s\neq r}}^{M}\dfrac{u_r-u_s-\e_1}{u_r-u_s+\e_1},\quad \Xi(z)=\prod_{\ell=1}^{N}(z-\xi_\ell),
\end{equation}
and the number of Bethe roots is $M=\sum_\ell n_\ell$. These equations resemble the Bethe equations of an inhomogeneous $sl(2)$ XXX spin chain with a twist parameter $q$.\footnote{In fact, in the superconformal case $\tilde{N}=2N$, it exactly reproduces the inhomogeneous XXX spin chain for an appropriate choice of masses $m_f$.} The TQ-equation associated to this system of Bethe roots reads
\begin{equation}\label{TQ-cutoff}
T(z)Q(z)=\Xi(z)\Xi(z+\e_1)Q(z+\e_1)+qm(z)Q(z-\e_1),
\end{equation}
Introducing the Q-polynomial as in \ref{def_q}, it is indeed possible to show that the RHS is a polynomial of degree $M+2N$, with $M$ zeros at $z=u_r$ as a consequence of the Bethe equations \ref{Bethe}. The TQ-equation \ref{def_q} is reproduced by further sending the number of Bethe roots $M$ to infinity, together with the cut-offs $\xi_\ell$ after a proper rescaling of the T and Q polynomials. More details on this limit will be provided in the work \cite{BF2015} to appear.

The NS limit of the expectation value \ref{avr_mass} of operators is also dominated by the single state $|\vec a,\vec Y^\ast\rangle$, and diagonal operators in the basis $|\vec a,\vec Y\rangle$ can be identified with their eigenvalues:\footnote{This is true for well-behaved operators for which the insertion does not modify significantly the saddle point equations.}
\begin{equation}
\la \Op\ra\sim  \la\vec a,\vec Y^\ast|\Op|\vec a,\vec Y^\ast\ra
\end{equation}
due to the simplification $\Zi\sim\Zv(\vec a,\vec Y^\ast)\Zf(\vec m;\vec a,\vec Y^\ast)$. From the action \ref{prop_CY} of the operator $\CY(z)$ on states $\aY$, replacing the coordinate $\phi_x$ of boxes that can be added to /removed from $\vec Y^\ast$ by Bethe roots, we find
\begin{equation}
\la\CY(z)\ra\sim \la\vec a,\vec Y^\ast|\CY(z)|\vec a,\vec Y^\ast\ra\sim\dfrac{Q(z)\Xi(z)}{Q(z-\e_1)},\quad \la\dfrac1{\CY(z)}\ra\sim \la\vec a,\vec Y^\ast|\dfrac1{\CY(z)}|\vec a,\vec Y^\ast\ra\sim\dfrac{Q(z-\e_1)}{Q(z)\Xi(z)}.
\end{equation}
Denoting $\chi_\text{NS}(z)$ the limit of the qq-character $\chi(z)$, the identity \ref{def_chi1} reproduces the TQ-equation \ref{TQ-cutoff} with the T-polynomial $T(z)=\chi_\text{NS}(z)\Xi(z)$ (the presence of the extra cut-off factor $\Xi(z)$ will be explained in \cite{BF2015}).

Our observation can be easily generalized to apply to linear quiver gauge theories. The NS limit for the $A_2$ quiver has been performed in \cite{Bourgine2014a}. Using the same procedure, the qq-character identity \ref{A2_qq_bis} reproduces the TQ-relation for an inhomogeneous $sl(3)$ XXX spin chain characterized by two sets of Bethe roots (equation (7.15) of \cite{Nekrasov2013}).

\section{Summary and concluding remarks}
In this paper, we developed a holomorphic field representation of SH$^c$ algebra. It has the merit to express the commutation relations and the finite rank representations of the operators in a compact form. Instanton partitions for $A_Q$ and $A^{(1)}_Q$-type quiver
gauge theories can be expressed concisely in terms of Gaiotto state, intertwiner operator, and a newly introduced flavor vertex operator that insert the contribution of fundamental hypermultiplets. A new characterization of the Gaiotto state and intertwiner has been established using the adjoint action of SH$^c$ holomorphic fields. It provided the infinite set of constraint on the instanton partition function from the chiral ring generating function proposed in \cite{Nekrasov2013}. These constraints are summarized in simple algebraic relations which were referred as the qq-character.

The qq-characters describe a quantum version of the Seiberg-Witten geometry \cite{Nekrasov:2012xe} in the NS limit $\e_2\to0$. In this setup, the $\e_2$-deformation introduces a form of second quantization of the TQ-system in which the T-polynomial is replaced by an operator acting in the Hilbert space of the rank $N$ representation. This should be compared with the recent results obtained in \cite{Bourgine2015a,Bourgine2015}, where the subleading corrections to the NS limit have been derived. These corrections are compatible (up to a quantum correction) with the second quantization of the NS action that is interpreted as the Yang-Yang functional of the underlying quantum integrable system. By comparing these two different approaches, a unified interpretation for the $\e_2$-deformation of quantum integrable systems should emerge.

There are some obvious generalizations of the current study for future work. In \cite{Nekrasov2013}, similar algebraic relations for the chiral generating functional were proposed for ADE type quiver gauge theories. In order to describe the bifurcation in
the quiver diagram, we need to find a SH$^c$ description of %the tri-fundamental representations. 
trivalent vertex which takes of the form:
\begin{eqnarray}
|\vec b\rangle\!\rangle_{\alpha\beta\gamma}=\sum_{\vec W} \Zv(\vec b, \vec W)^{-1/2} 
|\vec b, \vec W\rangle_\alpha\otimes |\vec b, \vec W\rangle_\beta\otimes |\vec b, \vec W\rangle_\gamma.
\end{eqnarray}
Here we added extra labels $\alpha,\cdots$ to specify the Hilbert spaces.
With the help of such operator, one may give the partition for the
$D_4$ quiver partition function, for example, as
\begin{eqnarray}
&&\otimes_{i=1}^3 \left( {}_{\alpha_i}\langle G,\vec a_i|q_i^{D^{(i)}_{0,1}}V_{\alpha_i\beta_i}(\vec a_i,\vec b|m_{i})\right)\cdot\   q_4^{D_{0,1}^{(1)}}\ |\vec b\rangle\!\rangle_{\beta_1\beta_2\beta_3}\nonumber\\
&&=\sum_{\vec Y_1,\vec Y_2, \vec Y_3, \vec W}q_4^{|\vec W|}\Zv(\vec b, \vec W)
\prod_{i=1}^{3} q_i^{|\vec Y_i|}\Zv(\vec a_i, \vec Y_i)\Zbf(\vec a_i,\vec Y_i; \vec{b}, \vec W|
m_i)\,.
\end{eqnarray}
In order to derive the qq-character for such extended cases, we need to
 find an analog of 
(\refOld{action_vertex1}, \refOld{action_vertex2}) for the
trivalent vertex.  At this moment, however, this seems not so simple
and we would like to leave it for future study.
%in terms of SH$^c$.
%Judging from the simplicity of the description of the bifundamental hypermultiplet, this objective may be
%accessible in a near future.  
Another possible direction proposed in \cite{Nekrasov2013} is the 5D version of the current analysis. It corresponds to the algebra studied by many authors in \cite{ding1997generalization, miki2007q, Feigin:2010qea,feigin2011quantum}
and has implications in 4D \cite{itoyama20132d,belavin2013bases}.
Since the building blocks are already known (for example, \cite{awata2010five, awata2011notes}), it would not be so difficult to perform a similar analysis in such set-up. In addition, our formulation of SH$^c$ seems particularly suited to the generalization to the six-dimensional $\Omega$-background $\mathbb{R}_{\e_1}^2\times\mathbb{R}_{\e_2}^2\times\mathbb{R}_{\e_3}^2$ in which the instanton partition function of $\mathcal{N}=2$ theories are expressed as a sum over plane partitions \cite{Szabo2015}.

In a different perspective, it is important to clarify the relations with integrable models. On one hand, a proposal of NPS in \cite{Nekrasov2013} suggests a connection between quantum geometry and the representation of the Yangian associated to the quiver Dynkin diagram. On the other hand, Maulik and Okounkov have proposed in \cite{Maulik:2012wi} the expression of the Yangian of $\widehat{gl}(1)$. In their formalism, the Dynkin diagram is regarded as the finite lattice of a spin system, where the spin degree of freedom is actually described by a free boson Fock space. In \cite{zhu2015yangian}, a short summary of \cite{Maulik:2012wi} and a possible supersymmetric generalization were presented.  While the two approaches are very different, the coproduct defined by the authors of \cite{Maulik:2012wi} coincides with the one employed in \cite{schiffmann2013cherednik}. Our analysis which relates the NPS qq-character \cite{Nekrasov2013} to the SH$^c$ algebra \cite{schiffmann2013cherednik} could provide an interesting link between the two Yangians.

\medskip
\textbf{Note added} After the first version of this paper was submitted to arXiv, N. Nekrasov published a paper \cite{Nekrasov2015a} where he studied Dyson-Schwinger equations for the instanton partition functions. He evaluated the effect of adding point-like instantons, and express this effect by an operator $\CY$ which is identical to ours. There seems to be a direct relation with our analysis and we hope to provide a more detailed comparison in the near future.

\medskip
\textbf{Acknowledgements}
J.-E. Bourgine thanks I.N.F.N. for his post-doctoral fellowship within the grant GAST, which has also partially supported this project, together with the UniTo-SanPaolo research  grant Nr TO-Call3-2012-0088 {\it ``Modern Applications of String Theory'' (MAST)}, the ESF Network {\it ``Holographic methods for strongly coupled systems'' (HoloGrav)} (09-RNP-092 (PESC)) and the MPNS--COST Action MP1210.
YM would like to thank Satoshi Nakamura and R.-D. Zhu for their comments
and discussions.
He is partially supported by Grants-in-Aid for Scientific Research (Kakenhi \#25400246) from MEXT, Japan.
HZ thanks Chaiho Rim for instructions and comments. He is supported by the National Research Foundation of Korea(NRF) grant funded by the Korea government(MSIP) (NRF-2014R1A2A2A01004951).

\appendix
\section{Comments on the notations}\label{app:conv}
In this paper, in order to ease the comparison
with the gauge theory, we use the omega background
parameters $\e_1,\e_2$ instead of the
CFT parameter $\beta=-\e_1/\e_2$ in \cite{Kanno2013, Matsuo2014}.
Since some results of these papers are used here,
we summarize the correspondence between the notations in this appendix.
As we use two parameters instead of one, we need some rescaling
and shift of parameters to compare with the results there. Adding a tilde to the notations in  \cite{Kanno2013, Matsuo2014}, the comparison goes as follows:
\ba
&& D_{0,n+1}=(\e_2)^n \tilde D_{0,n+1},\quad D_{\pm 1,n}=(\e_2)^n \tilde D_{\pm 1,n},\quad
E_{n}=(\e_2)^n \tilde E_n\label{rescale}\,,\\
&& a_\ell=-\e_2 \tilde a_\ell+\e_+, \quad z=\e_2/\tilde\zeta,
\quad c_n=(-\e_2)^n \tilde c_n,\\
&& \phi(x)|_{x\in A(Y_\ell)} =-\e_2(\tilde a_\ell+\tilde A_t(Y_\ell)),\quad
\phi(x)|_{x\in R(Y_\ell)} =-\e_2(\tilde a_\ell+\tilde B_t(Y_\ell)).
\ea
We note that under the rescaling \ref{rescale}, the algebra
(\refOld{SH1}--\refOld{SH4}) remains the same.

%\section{Characterization of Gaiotto states and intertwiner under SH$^c$ transformations}
\section{Proof of the recursion formulae for Gaiotto states
and intertwiner}\label{AppA}
Since the Gaiotto state can be derived from the intertwiner
\ref{I2G}, it will be sufficient to prove 
(\refOld{action_vertex1}, \refOld{action_vertex2}).
We need a few formulae to characterize the behavior of the 
instanton partition function building blocks (here bifundamental 
and the vector contributions) under the variations of the number of boxes in the $N$-tuple $\vec Y$.
\begin{eqnarray}
\frac{\Zbf(\vec a, \vec Y+x;\vec b, \vec W|m)}
{\Zbf(\vec a, \vec Y;\vec b, \vec W|m)}&=&
\frac{\prod_{y\in A(\vec W)} 
(\phi_x-\phi_y+\e_+ -m)}{\prod_{y\in R(\vec W)} (\phi_x-\phi_y-m)},
\label{prop_Zbf1}\\
\frac{\Zbf(\vec a, \vec Y-x;\vec b, \vec W|m)}
{\Zbf(\vec a, \vec Y;\vec b, \vec W|m)}&=&
\frac{\prod_{y\in R(\vec W)} 
(\phi_x-\phi_y-m)}{\prod_{y\in A(\vec W)} (\phi_x-\phi_y+\e_+ -m)},
\label{prop_Zbf2}\\
\frac{\Zbf(\vec a, \vec Y;\vec b, \vec W+x|m)}
{\Zbf(\vec a, \vec Y;\vec b, \vec W|m)}&=&
\frac{\prod_{y\in A(\vec Y)} 
(\phi_x-\phi_y+m)}{\prod_{y\in R(\vec Y)} (\phi_x-\phi_y+m-\e_+)},
\label{prop_Zbf3}\\
\frac{\Zbf(\vec a, \vec Y;\vec b, \vec W-x|m)}
{\Zbf(\vec a, \vec Y;\vec b, \vec W|m)}&=&
\frac{\prod_{y\in R(\vec Y)} 
(\phi_x-\phi_y+m-\e_+)}{\prod_{y\in A(\vec Y)} (\phi_x-\phi_y+m)},
\label{prop_Zbf4}
\end{eqnarray}
\begin{eqnarray}
%\begin{equation}\label{prop_Zv}
\dfrac{\Zv(\vec a,\vec Y+x)}{\Zv(\vec a,\vec Y)}&=&-\dfrac1{\e_1\e_2}\dfrac{\prod_{y\in R(\vec Y)}(\phi_x-\phi_y)(\phi_x-\phi_y-\e_+)}{\prod_{\superp{y\in A(\vec Y)}{y\neq x}}(\phi_x-\phi_y)(\phi_x-\phi_y+\e_+)},\label{prop_Zv}\\
\quad \dfrac{\Zv(\vec a,\vec Y-x)}{\Zv(\vec a,\vec Y)}&=&-\dfrac1{\e_1\e_2}\dfrac{\prod_{y\in A(\vec Y)}(\phi_x-\phi_y)(\phi_x-\phi_y+\e_+)}{\prod_{\superp{y\in R(\vec Y)}{y\neq x}}(\phi_x-\phi_y)(\phi_x-\phi_y-\e_+)}\,.\label{prop_Zv2}
%\end{equation}
\end{eqnarray}
These formulae were used in \cite{Kanno2013} to prove
the recursive properties of the quiver gauge theories.
Essentially the same computation shows up here.
We evaluate the action of $D_\eta(z)$ on the intertwiner:
\begin{equation}
V(\vec a,\vec b|m)=\sum_{\vec{Y}, \vec{W}}
\bZbf(\vec a,\vec Y;\vec b,\vec W|m)|\vec{a},\vec Y\rangle
\langle \vec b,\vec W|\,,
\end{equation}
with $\bZbf(\vec a,\vec Y;\vec b,\vec W|m):=\sqrt{\Zv(\vec a,\vec Y)\Zv(\vec b,\vec W)}\Zbf(\vec a, \vec Y:\vec b, \vec W|m)$.
We first evaluate the action of $D_{-1}(z)$
on the intertwiner from the left.  
It is easily deduced from its action on states $\aY$,
\begin{equation}
D^{\vec a}_{-1}(z)V(\vec a,\vec b|m)=\sum_{\vec Y,\vec W}\bZbf(\vec a,\vec Y;\vec b,\vec W|m) \sum_{x\in R(\vec Y)}\dfrac{\L_x(\vec Y)}{z-\phi_x}
|\vec a,\vec Y-x\rangle\langle \vec b,\vec W|.
\end{equation}
An alternative expression can be obtained after noticing that the inverse images of a state $\aY$ under the mapping $D_{-1}(z)$ are the states $|\vec a,\vec Y+x\rangle$ for $x\in A(\vec Y)$ because the action of the operator removes one box. Since the states $\aY$ form a basis of the vector space $\CV_{\vec a}$, it is possible to write
\begin{equation}
D^{\vec a}_{-1}(z)V(\vec a,\vec b|m)=\sum_{\vec Y,\vec W}\sum_{x\in A(\vec Y)}\dfrac{\L_x(\vec Y+x)}{z-\phi_x}\bZbf(\vec a,\vec Y+x;\vec b,\vec W|m)
|\vec a,\vec Y\rangle\langle \vec b,\vec W|.
\end{equation}
It is convenient to rewrite this expression as follows, using the fact that $\L_x(\vec Y+x)^2=\L_x(\vec Y)^2$, $\forall x\in A(\vec Y)$:\footnote{In this expression, and the analysis hereafter, 
the correct choice of sign is verified by comparing with the direct action of SH$^c$ generators on states with a small number of boxes.}
\begin{equation}\label{leftact-1}
D^{\vec a}_{-1}(z)V(\vec a,\vec b|m)=\sum_{\vec Y,\vec W}\sum_{x\in A(\vec Y)}\dfrac{\L_x(\vec Y)}{z-\phi_x}
\frac{\bZbf(\vec a,\vec Y+x;\vec b,\vec W|m)}{\bZbf(\vec a,\vec Y;\vec b,\vec W|m)} 
\bZbf(\vec a,\vec Y;\vec b,\vec W|m)
|\vec a,\vec Y\rangle\langle \vec b,\vec W|.
\end{equation}
We evaluate the ratio for $\bZbf$ from \ref{prop_Zbf1} and \ref{prop_Zv}
and put the explicit form of $\Lambda_x(\vec Y)$ \ref{prop_L}:
\begin{equation}
\L_x(\vec Y)
\frac{\bZbf(\vec a,\vec Y+x;\vec b,\vec W|m)}{\bZbf(\vec a,\vec Y;\vec b,\vec W|m)}
=\frac{1}{\sqrt{-\e_1\e_2}}\frac{\prod_{y\in R(\vec Y)}(\phi_x-\phi_y-\e_+)}{\prod_{\superp{y\in A(\vec Y)}{y\neq x}}(\phi_x-\phi_y)}
\frac{\prod_{y\in A(\vec W)}(\phi_x-\phi_y-m+\e_+)}{\prod_{y\in R(\vec W)}(\phi_x-\phi_y-m)}\,.
\end{equation}
The action of $D_{-1}$ from the right can be evaluated similarly,
\begin{equation}\label{rightact-1}
-V(\vec a,\vec b|m)D^{\vec b}_{-1}(z')=\sum_{\vec Y,\vec W}\sum_{x\in R(\vec W)}\dfrac{\L_x(\vec W)}{z'-\phi_x}
\frac{\bZbf(\vec a,\vec Y;\vec b,\vec W-x|m)}{\bZbf(\vec a,\vec Y;\vec b,\vec W|m)} 
\bZbf(\vec a,\vec Y;\vec b,\vec W|m)
|\vec a,\vec Y\rangle\langle \vec b,\vec W|.
\end{equation}
From \ref{prop_Zbf4} and \ref{prop_Zv2}, the factor in the middle
takes the form:
\begin{equation}
\L_x(\vec W)
\frac{\bZbf(\vec a,\vec Y;\vec b,\vec W-x|m)}{\bZbf(\vec a,\vec Y;\vec b,\vec W|m)}
=\frac{1}{\sqrt{-\e_1\e_2}}\frac{\prod_{y\in A(\vec W)}(\phi_x-\phi_y+\e_+)}{\prod_{\superp{y\in R(\vec W)}{y\neq x}}(\phi_x-\phi_y)}
\frac{\prod_{y\in R(\vec Y)}(\phi_x-\phi_y+m-\e_+)}{\prod_{y\in A(\vec Y)}(\phi_x-\phi_y+m)}\,.
\end{equation}
To add \ref{leftact-1} and \ref{rightact-1}, we use the following identity
to simplify the formula (we put $z'=z-m$)
\begin{eqnarray}
&&\sum_{x\in A(\vec Y)}\frac{1}{z-\phi_x}\frac{\prod_{y\in R(\vec Y)}(\phi_x-\phi_y-\e_+)}{\prod_{\superp{y\in A(\vec Y)}{y\neq x}}(\phi_x-\phi_y)}
\frac{\prod_{y\in A(\vec W)}(\phi_x-\phi_y-m+\e_+)}{\prod_{y\in R(\vec W)}(\phi_x-\phi_y-m)}
\nonumber\\
&&+\sum_{x\in R(\vec W)}\frac{1}{z-m-\phi_x}
\frac{\prod_{y\in A(\vec W)}(\phi_x-\phi_y+\e_+)}{\prod_{\superp{y\in R(\vec W)}{y\neq x}}(\phi_x-\phi_y)}
\frac{\prod_{y\in R(\vec Y)}(\phi_x-\phi_y+m-\e_+)}{\prod_{y\in A(\vec Y)}(\phi_x-\phi_y+m)}\nonumber\\
&&~~~~~=\mathrm{P}^-_z\left(\frac{\prod_{y\in R(\vec Y)}(z-\phi_y-\e_+)}{\prod_{y\in A(\vec Y)}(z-\phi_y)}
\frac{\prod_{y\in A(\vec W)}(z-\phi_y-m+\e_+)}{\prod_{y\in R(\vec W)}(z-\phi_y-m)}\right).
\end{eqnarray}
This formula is obtained by comparing the residue of
the poles on both sides.  Finally, we note that 
by using \ref{prop_CY}, we obtain
\begin{eqnarray}
&&\mathrm{P}^-_z\left(\frac{\prod_{y\in R(\vec Y)}(z-\phi_y-\e_+)}{\prod_{y\in A(\vec Y)}(z-\phi_y)}
\frac{\prod_{y\in A(\vec W)}(z-\phi_y-m+\e_+)}{\prod_{y\in R(\vec W)}(z-\phi_y-m)}\right)|\vec a,\vec Y\rangle\langle \vec b,\vec W|\nonumber\\
&&~~~~=
\mathrm{P}^-_z\left(\frac{1}{\CY^{(1)}(z)}|\vec a,\vec Y\rangle\langle \vec b,\vec W| \CY^{(2)}(z-m+\e_+)\right).
\end{eqnarray}
After combining everything, we prove \ref{action_vertex1}. 
The proof of \ref{action_vertex2} is completely parallel 
and we omit it here.

\section{Some calculations of commutation relations in the rank $N$ representation}\label{sec:commutator}
For completeness, we present here some explicit computations for the commutators of holomorphic generators. 

\paragraph{$[D_0(z), D_1(w)]$}
From \ref{act_D}, 
\begin{eqnarray}
D_1(w)D_0(z)|\vec{a}, \vec Y\rangle &=& \sum_{x\in \vec Y} \frac{1}{z-\phi_x}
\sum_{y\in A(\vec Y)}\frac{\L_y(\vec Y)}{w-\phi_y}|\vec a,\vec Y+y\rangle,\\
D_0(z) D_1(w)|\vec{a}, \vec Y\rangle &=&
\sum_{y\in A(\vec Y)} \sum_{x\in \vec Y+y} \frac{1}{z-\phi_x}\frac{\L_y(\vec Y)}{w-\phi_y}|\vec a,\vec Y+y\rangle.
\end{eqnarray}
The difference is the inclusion of the box $y$ in $\vec Y+y$. So we obtain
the first relation in \ref{comm_SHc}.
\begin{eqnarray}
\left[D_0(z), D_1(w)\right]|\vec a, \vec Y\rangle &=&
\sum_{y\in A(\vec Y)} \frac{\L_y(\vec Y)}{(z-\phi_y)(w-\phi_y)}|\vec a, \vec Y\rangle
=\sum_{y\in A(\vec Y)} \frac{1}{z-w}\left(\frac{\L_y(\vec Y)}{w-\phi_y}-
\frac{\L_y(\vec Y)}{z-\phi_y}\right)|\vec a, \vec Y\rangle\nonumber\\
&=& \frac{D_1(w)-D_1(z)}{z-w}|\vec a,\vec Y\rangle.
\end{eqnarray}

\paragraph{$[D_1(z), D_{-1}(w)]$}
\begin{eqnarray}
D_1(z)D_{-1}(w)|\vec a, \vec Y\rangle &=&\sum_{x\in R(\vec Y)}\sum_{y\in A(\vec Y-x)}
\frac{\L_x(\vec Y)}{w-\phi_x}\frac{\L_y(\vec Y-x)}{z-\phi_y}|\vec a,\vec Y-x+y\rangle
\nonumber\\
&=& \sum_{x\in R(\vec Y)}\sum_{y\in A(\vec Y)}
\frac{\L_x(\vec Y)}{w-\phi_x}\frac{\L_y(\vec Y-x)}{z-\phi_y}|\vec a,\vec Y-x+y\rangle
+\sum_{x\in R(\vec Y)}
\frac{\L_x(\vec Y)}{w-\phi_x}\frac{\L_x(\vec Y-x)}{z-\phi_x}|\vec a,\vec Y\rangle,
\nonumber
\end{eqnarray}
where we have assumed for $\vec Y$ a generic form such that $A(\vec Y-x)=A(\vec Y)\cup\{x\}$. From the explicit form of $\L_x(\vec Y)$, one may prove for $y\in A(\vec Y)$:
\begin{equation}
\L_y(\vec Y-x)^2=\L_y(\vec Y)^2\frac{S(\phi_x-\phi_y)}{S(\phi_y-\phi_x)},\quad
\L_x(\vec Y-x)^2=\L_x(\vec Y)^2\,,
\end{equation}
where $S(z)$ 
is the scattering factor which appeared in \ref{braiding}.
After combining them, $D_1(z)D_{-1}(w)|\vec a, \vec Y\rangle$ becomes,
\begin{equation}
\sum_{x\in R(\vec Y)}\sum_{y\in A(\vec Y)}
\frac{\L_x(\vec Y)}{w-\phi_x}\frac{\L_y(\vec Y)}{z-\phi_y}\left(
\frac{S(\phi_x-\phi_y)}{S(\phi_y-\phi_x)}
\right)^{1/2}|\vec a,\vec Y-x+y\rangle
+\sum_{x\in R(\vec Y)}
\frac{(\L_x(\vec Y))^2}{(w-\phi_x)(z-\phi_x)}|\vec a,\vec Y\rangle\,.
\end{equation}
$D_{-1}(w)D_1(z)|\vec a, \vec Y\rangle$ is evaluated similarly, assuming that $R(\vec Y+x)=R(\vec Y)$,
\begin{eqnarray}
&& \sum_{y\in A(\vec Y)}\sum_{x\in R(\vec Y)}
\frac{\L_x(\vec Y)}{w-\phi_x}\frac{\L_y(\vec Y)}{z-\phi_y}\left(
\frac{S(\phi_x-\phi_y)}{S(\phi_y-\phi_x)}
\right)^{1/2}|\vec a,\vec Y-x+y\rangle
+\sum_{x\in A(\vec Y)}
\frac{(\L_x(\vec Y))^2}{(w-\phi_x)(z-\phi_x)}|\vec a,\vec Y\rangle\,.
\end{eqnarray}
In $\left[ D_1(z), D_{-1}(w)\right]$, the first term cancels
and the second term gives $\frac{E(z)-E(w)}{\e_+(z-w)}|\vec a, \vec Y\rangle$ after the
use of the relations \ref{L2}, \ref{prop_L} and \ref{Ez}. Degenerate situations should be evaluated case by case, but the general conclusion remains unchanged.

%%%%%% Removing the D1D1 part
% \paragraph{$D_1(z)D_1(w)$ and the braiding relation}
% Acting twice with the holomorphic operator $D_1(z)$,
% \begin{eqnarray}
% D_1(z)D_1(w)|\vec a,\vec Y\rangle &=&
% \sum_{x\in A(\vec Y)}\sum_{y\in A(\vec Y+x)} \frac{\L_x(\vec Y)}{w-\phi_x}
% \frac{\L_y(\vec Y+x)}{z-\phi_y}|\vec a,\vec Y+x+y\rangle\nonumber\\
% &=& \sum_{\superp{x,y\in A(\vec Y)}{x\neq y}} \frac{\L_x(\vec Y)}{w-\phi_x}
% \frac{\L_y(\vec Y)}{z-\phi_y}
% \left( \frac{S(\phi_y-\phi_x)}{S(\phi_x-\phi_y)}\right)^{1/2}|\vec a,\vec Y+x+y\rangle\nonumber\\
% &&+ \sum_{x\in A(\vec Y)}\sum_{a=1,2} \frac{\L_x(\vec Y)}{w-\phi_x}
% \frac{\L_{x+e_a}(\vec Y+x)}{z-\phi_x-\e_a}|\vec a,\vec Y+x+(x+\e_a)\rangle\,.
% \end{eqnarray}
% To establishe the second equality, the relation $\L_y(\vec Y+x)^2=\L_y(\vec Y)^2 S(\phi_y-\phi_x)/S(\phi_x-\phi_y)$ has been employed. 
% The contributions in the third line does not exist
% when $\vec Y+x+(x+\e_a)$ is not a Young diagram.
% It is known that the product of the form $D_\eta(z) D_\eta(w)$ ($\eta=\pm1$) enjoys the braiding
% relation, \cite{arbesfeld2012presentation},
% \begin{equation}
% S(z-w)^{-\eta}D_\eta(z)D_\eta(w)\sim S(w-z)^{-\eta}D_\eta(w)D_\eta(z)\,.
% \end{equation}
% We note that this relation holds only approximately. Namely 
% both hand side has the same behavior near
% the poles at $z=\phi_x, w=\phi_y$, $x, y\in A/R(\vec Y)$, $x\neq y$. However, the behavior is different at some particular poles, such as $w=\phi_x$ and $z=\phi_x+\e_a$.

%%%%%%%%%%%%%%%%%%%%%%%%%%%%%%
\section{Detailed analysis of the qq-characters}\label{AppB}
\subsection{Matter case}
There are different possibilities for the insertion position of the SH$^c$ operator $D_{-1}(z)$, the simplest option is to insert it on the left of the mass operator as in $\langle G,\vec a|q^D D_{-1}(z)U(\vec m)\Ga$, and then use the braiding relation \ref{DU} to move it to the right. The two formulas \ref{action_Gaiotto} and \ref{action_Gaiotto_adj} for the action of $D_{-1}(z)$ provides the identity \ref{chi_mass_pm}. However, this is not the unique choice for the insertion of the $D_{-1}(z)$, a second possibility is to insert it on the right of the mass operator,
\begin{equation}
\langle G,\vec a|q^D U(\vec m)D_{-1}(z)\Ga.
\end{equation}
A new braiding relation is needed in order to move the SH$^c$ operator to the left, which is deduced from \ref{com_D_Phi}:
\begin{equation}
U(\vec m)D_{-1}(z)=\left(\dfrac{D_{-1}(z)}{m(z)}-\sum_{f=1}^{\tilde{N}}\dfrac{D_{-1}(\mf)}{z-\mf}\prod_{\superp{f'=1}{f'\neq f}}^{\tilde{N}}\dfrac1{\mf-m^{(f')}}\right)U(\vec m).
\end{equation}
The action of $D_{-1}(z)$ on the Gaiotto state is then computed from \ref{action_Gaiotto} and \ref{action_Gaiotto2}, leading to the identity
\begin{equation}
\dfrac1{m(z)}\mathrm{P}_z^-\la\CY(z+\e_+)\ra-\sum_{f=1}^{\tilde{N}}\dfrac1{z-\mf}\prod_{\superp{f'=1}{f'\neq f}}^{\tilde{N}}\dfrac1{\mf-m^{(f')}}\la\CY(\mf+\e_+)-\Pi(\mf)\ra=-q\la\dfrac{1}{\CY(z)}\ra,
\end{equation}
where the operator $\Pi(z)=\mathrm{P}_z^+\CY(z+\e_+)$ has been introduced to simplify the expression. As a result, the fundamental qq-character defined in \ref{def_chi1} obeys
\begin{equation}
\chi(z)=\la\Pi(z)\ra+m(z)\sum_{f=1}^{\tilde{N}}\dfrac1{z-\mf}\prod_{\superp{f'=1}{f'\neq f}}^{\tilde{N}}\dfrac1{\mf-m^{(f')}}\la\CY(\mf+\e_+)-\Pi(\mf)\ra\,.
\end{equation}
In the last term, the apparent poles are cancelled by the zeros of $m(z)$ and the RHS is a polynomial. The compatibility with the identity \ref{chi_mass_pm} implies the following equality,
\begin{equation}
m(z)\sum_{f=1}^{\tilde{N}}\dfrac1{z-\mf}\prod_{\superp{f'=1}{f'\neq f}}^{\tilde{N}}\dfrac1{\mf-m^{(f')}}\la\CY(\mf+\e_+)-\Pi(\mf)\ra=\mathrm{P}_z^+q\la\dfrac{m(z)}{\CY(z)}\ra.
\end{equation}

\subsection{Second qq-character of the $A_1$ quiver}
\subsubsection{Double action of SH$^c$ operators $D_\eta(z)$}
As a preliminary step it is necessary to derive the action of two operators $D_{-1}(z_1)$ and $D_1(z_2)$ with different arguments on a Gaiotto state $\Ga$. The method is the same as in the case of a single operator, and upon using the property
\begin{equation}
\langle\vec a,\vec Y+x|\CY(z+\e_+)|\vec a,\vec Y+x\rangle=S(z-\phi_x)\langle\vec a,\vec Y|\CY(z+\e_+)\aY,
\end{equation}
with $S(z)$ the scattering factor defined in \ref{braiding}, it is possible to write down
{\small
\begin{align}
\begin{split}
D_{-1}(z_1)D_{1}(z_2)\Ga=\dfrac{1}{\e_1\e_2}\sum_{\vec Y}\sqrt{\Zv(\vec a,\vec Y)}\sum_{x\in A(\vec Y)}&\dfrac1{z_1-\phi_x}\dfrac{\prod_{y\in R(\vec Y)}(\phi_{xy}-\e_+)}{\prod_{\superp{y\in A(\vec Y)}{y\neq x}}\phi_{xy}}\mathrm{P}_{z_2}^-S(z_2-\phi_x)\CY(z_2+\e_+)\aY,
\end{split}
\end{align}}
where the sign ambiguity has been fixed by comparing the coefficient of the vacuum state with the direct action of the SH$^c$ generators. Inserting the pole decomposition of 
\begin{equation}
\dfrac{S(z_2-\phi_x)}{z_1-\phi_x}=\dfrac{S(z_{21})}{z_1-\phi_x}+\dfrac{\e_1\e_2}{\e_+z_{12}}\dfrac1{z_2-\phi_x}-\dfrac{\e_1\e_2}{\e_+(z_{12}-\e_+)}\dfrac1{z_2-\phi_x+\e_+},
\end{equation}
with the shortcut notation $z_{21}=z_2-z_1$, it is possible to perform the summation over $x\in A(\vec Y)$:
\begin{align}
\begin{split}
&D_{-1}(z_1)D_{1}(z_2)\Ga\\
&=\dfrac{1}{\e_1\e_2}\sum_{\vec Y}\sqrt{\Zv(\vec a,\vec Y)}\left[\mathrm{P}_{z_2}^-S(z_{21})\dfrac{\CY(z_2+\e_+)}{\CY(z_1)}+\dfrac{\e_1\e_2}{\e_+z_{12}}\dfrac{\CY(z_2+\e_+)}{\CY(z_2)}-\dfrac{\e_1\e_2}{\e_+(z_{12}-\e_+)}\right]\aY,
\end{split}
\end{align}
where it has been used that the last two terms in the brackets have no polynomial part at infinity, and consequently in the last term the two factors $\CY(z_2+\e_+)$ have cancelled each other. This result can be written in the compact form
\begin{equation}\label{DD_Ga}
D_{-1}(z_1)D_{1}(z_2)\Ga=\left[\dfrac{1}{\e_1\e_2}\mathrm{P}_{z_2}^-S(z_{21})\dfrac{\CY(z_2+\e_+)}{\CY(z_1)}+\dfrac1{\e_+z_{12}}\dfrac{\CY(z_2+\e_+)}{\CY(z_2)}-\dfrac1{\e_+(z_{12}-\e_+)}\right]\Ga.
\end{equation}

The action of the commuted operators $D_{1}(z_2)D_{-1}(z_1)$ is derived from the same method,
\begin{equation}
D_{1}(z_2)D_{-1}(z_1)\Ga=\dfrac1{\CY(z_1)}\left[\dfrac{S(z_{21})}{\e_1\e_2}\mathrm{P}_{z_2}^-\CY(z_2+\e_+)-\dfrac{\mathrm{P}_{z_1}^-\CY(z_1+\e_+)}{\e_+ z_{21}}+\dfrac{\mathrm{P}_{z_1}^-\CY(z_1)}{\e_+(z_{21}+\e_+)}\right]\Ga.
\end{equation}
This expression is simplified employing the following identity,
\begin{equation}
\mathrm{P}_{z_2}^-\left[S(z_{21})\CY(z_2+\e_+)\right]=S(z_{21})\mathrm{P}_{z_2}^-\CY(z_2+\e_+)+\dfrac{\e_1\e_2}{\e_+ z_{21}}\mathrm{P}_{z_1}^+\CY(z_1+\e_+)-\dfrac{\e_1\e_2}{\e_+(z_{21}+\e_+)}\mathrm{P}_{z_1}^+\CY(z_1),
\end{equation}
obtained by decomposition of  $S(z_{21})$ as a sum over single poles and of $\CY(z_2+\e_+)$ into positive and negative powers. As a result, we find
\begin{equation}\label{DD_Ga2}
D_{1}(z_2)D_{-1}(z_1)\Ga=\left[\dfrac{1}{\e_1\e_2}\mathrm{P}_{z_2}^-S(z_{21})\dfrac{\CY(z_2+\e_+)}{\CY(z_1)}+\dfrac1{\e_+z_{12}}\dfrac{\CY(z_1+\e_+)}{\CY(z_1)}-\dfrac1{\e_+(z_{12}-\e_+)}\right]\Ga.
\end{equation}
Taking the difference between \ref{DD_Ga} and \ref{DD_Ga2}, the commutation relation \ref{comm_SHc} between $D_{-1}(z_1)$ and $D_{1}(z_2)$ is recovered, with the action of $E(z_\a)$ on Gaiotto states given in \ref{E_CY} by the ratio of $\CY$ operators with shifted arguments.

\subsubsection{Derivation of the second qq-character}
The expression of the second qq-character follows from the consideration of the symmetrized action \ref{DD_Ga} of two SH$^c$ operators inside two Gaiotto states,
\begin{equation}
\langle \vec G,\vec a|q^D\left[D_{-1}(z_1)\,D_{1}(z_2)+D_{-1}(z_2)\,D_{1}(z_1)\right]\Ga.
\end{equation}
This quantity can be computed either using the right action \ref{DD_Ga} of two operators on the Gaiotto state, or from the right \ref{action_Gaiotto}, \ref{action_Gaiotto2} and left \ref{action_Gaiotto_adj} actions of a single SH$^c$ operator. These two possible ways of calculation furnish the following identity:
{\small
\begin{align}
\begin{split}
&-\dfrac{2q^{-1}}{\e_1\e_2}\mathrm{P}_{z_1}^-\mathrm{P}_{z_2}^-\la\CY(z_1+\e_+)\CY(z_2+\e_+)\ra\\
&=\dfrac1{\e_+ z_{12}}\la\dfrac{\CY(z_2+\e_+)}{\CY(z_2)}-\dfrac{\CY(z_1+\e_+)}{\CY(z_1)}\ra+\dfrac1{\e_1\e_2}\mathrm{P}_{z_2}^-\left[S(z_{21})\la\dfrac{\CY(z_2+\e_+)}{\CY(z_1)}\ra\right]+\dfrac1{\e_1\e_2}\mathrm{P}_{z_1}^-\left[S(z_{12})\la\dfrac{\CY(z_1+\e_+)}{\CY(z_2)}\ra\right]-\dfrac{2}{z_{12}^2-\e_+^2}.
\end{split}
\end{align}
}
The second line involves the commutator of $D_{-1}(z_1)$ with $D_{1}(z_2)$ evaluated in the Gaiotto states average. The same quantity can also be computed by direct right \ref{action_Gaiotto}, \ref{action_Gaiotto2} and left \ref{action_Gaiotto_adj} actions on Gaiotto states, leading to a second identity:
\begin{equation}
\dfrac1{\e_+ z_{12}}\la\dfrac{\CY(z_2+\e_+)}{\CY(z_2)}-\dfrac{\CY(z_1+\e_+)}{\CY(z_1)}\ra=-\dfrac{q^{-1}}{\e_1\e_2}\mathrm{P}_{z_1}^-\mathrm{P}_{z_2}^-\la\CY(z_1+\e_+)\CY(z_2+\e_+)\ra+\dfrac{q}{\e_1\e_2}\la\dfrac1{\CY(z_1)\CY(z_2)}\ra.
\end{equation}
Replacing the average of the commutator in the first identity, and introducing the positive part $\Pi(z)=\mathrm{P}_z^{+}\CY(z+\e_+)$ produces
\begin{align}
\begin{split}\label{prop_chi2}
0=&q^{-1}\la(\CY(z_1+\e_+)-\Pi(z_1))(\CY(z_2+\e_+)-\Pi(z_2))\ra+\mathrm{P}_{z_1}^-\left[S(z_{12})\la\dfrac{\CY(z_1+\e_+)}{\CY(z_2)}\ra\right]\\
&+\mathrm{P}_{z_2}^-\left[S(z_{21})\la\dfrac{\CY(z_2+\e_+)}{\CY(z_1)}\ra\right]+q\la\dfrac1{\CY(z_1)\CY(z_2)}\ra-\dfrac{2\e_1\e_2}{z_{12}^2-\e_+^2}.
\end{split}
\end{align}
This relation implies for the second qq-character \ref{chi2},
\begin{equation}
\label{PZPZ}
\mathrm{P}_{z_1}^-\mathrm{P}_{z_2}^-\chi_2(z_1,z_2)=\dfrac{2q\e_1\e_2}{z_{12}^2-\e_+^2}.
\end{equation}
This is not enough to conclude on the polynomiality of the second qq-character, because of the presence of the cross-terms
\begin{equation}
\mathrm{P}_{z_1}^-\mathrm{P}_{z_2}^+\chi_2(z_1,z_2)=\la\Pi(z_2)\left(\CY(z_1+\e_+)-\Pi(z_1)+q\dfrac{S(z_{21})}{\CY(z_1)}\right)\ra-q\dfrac{\e_1\e_2}{\e_+}\la\dfrac{\Pi(z_1)}{z_{21}\CY(z_1)}-\dfrac{\Pi(z_1+\e_+)}{(z_{21}+\e_+)\CY(z_1)}\ra.
\end{equation}
On the other hand, we know the explicit expression of the second qq-character for small $N$, and can use it to deduce the expression of $\chi_2(z_1,z_2)$. First, it is noted that after the introduction of the orthogonal projector in \ref{prop_chi2}, $\chi_2$ can be rewritten as the average of an operator $\Pi_2(z_1,z_2)$
\begin{equation}
\chi_2(z_1,z_2)=\la \Pi_2(z_1,z_2)\ra,
\end{equation}
defined as
{\small
\begin{equation}
\Pi_2(z_1,z_2)=-\Pi(z_1)\Pi(z_2)+\mathrm{P}_{z_1}^+\left[\Pi(z_1)\left(\CY(z_2+\e_+)+q\dfrac{S(z_{12})}{\CY(z_2)}\right)\right]+\mathrm{P}_{z_2}^+\left[\Pi(z_2)\left(\CY(z_1+\e_+)+q\dfrac{S(z_{21})}{\CY(z_1)}\right)\right]+\dfrac{2q\e_1\e_2}{z_{12}^2-\e_+^2}.
\end{equation}
}
At first sight, it is not clear whether this quantity is a polynomial and we had to check it case by case using the explicit expression of the polynomial operator $\Pi(z)$ given in \ref{chi1_expl} for $N=1,2$.

\paragraph{Case $N=1$:} In this case $\Pi(z)$ is a scalar and can be taken out of the vacuum expectation values, i.e. $\la\Pi(z)\cdots\ra=\Pi(z)\la\cdots\ra$. Since it is a polynomial of degree one, it satisfies $\mathrm{P}_{z_1}^+S(z_{12})\Pi(z_1)=\Pi(z_1)$ and as a result
\begin{equation}
\la \Pi_2(z_1,z_2)\ra=-\Pi(z_1)\Pi(z_2)+\Pi(z_1)\chi(z_2)+\Pi(z_2)\chi(z_1)+q\dfrac{2\e_1\e_2}{z_{12}^2-\e_+^2}
\end{equation}
where $\chi(z)$ is the fundamental qq-character given in \ref{def_chi1}. In this simple case, $\chi(z)=\Pi(z)$ which provides the final result
\begin{equation}
\la \Pi_2(z_1,z_2)\ra=\chi(z_1)\chi(z_2)+q\dfrac{2\e_1\e_2}{z_{12}^2-\e_+^2}.
\end{equation}

\paragraph{Case $N=2$:} In this case, $\Pi(z)$ is a polynomial of degree two that satisfies $\mathrm{P}_{z_1}^+S(z_{12})\Pi(z_1)=\Pi(z_1)+\e_1\e_2$. It follows that 
\begin{align}
\begin{split}
\Pi_2(z_1,z_2)&=-\Pi(z_1)\Pi(z_2)+\Pi(z_1)\left(\CY(z_2+\e_+)+\dfrac{q}{\CY(z_2)}\right)+\e_1\e_2\dfrac{q}{\CY(z_2)}\\
&+\Pi(z_2)\left(\CY(z_1+\e_+)+\dfrac{q}{\CY(z_1)}\right)+\e_1\e_2\dfrac{q}{\CY(z_1)}+q\dfrac{2\e_1\e_2}{z_{12}^2-\e_+^2}.
\end{split}
\end{align}
The explicit expression of $\Pi(z)$ deduced from \ref{exp_CY} allows to show that 
\begin{equation}
\la\Pi(z_1)\left(\CY(z_2+\e_+)+\dfrac{q}{\CY(z_2)}\right)+\e_1\e_2\dfrac{q}{\CY(z_2)}\ra=\dfrac1{\Zi}D_{z_1}\left(\Zi\chi(z_2)\right)=\dfrac1{\Zi}D_{z_1}D_{z_2}\Zi.
\end{equation}
with the shifted derivative
\begin{equation}
D_{z_\a}=(z_\a+\e_+)^2-a^2+\e_1\e_2q\p_q,\quad \chi(z)=\la\Pi(z)\ra=\dfrac{D_z\CZ_\text{inst}}{\Zi}.
\end{equation}
Using this expression we arrive at 
\begin{equation}
\chi_2(z_1,z_2)=\la \Pi_2(z_1,z_2)\ra=\dfrac1{\Zi}\left[D_{z_1}D_{z_2}+2q\dfrac{\e_1\e_2}{z_{12}^2-\e_+^2}\right]\CZ_\text{inst},\quad \la\Pi(z_1)\Pi(z_2)\ra=\dfrac1{\Zi}D_{z_1}D_{z_2}\Zi\,,
\end{equation}
replacing $z_1=z+\nu_1$ and $z_2=z+\nu_2$, this quantity is obviously a polynomial in $z$.

%
%\bibliographystyle{utphys}
%\bibliography{./BMZ.bib}
%\bibliography{./fullbib.bib}
\providecommand{\href}[2]{#2}\begingroup\raggedright\endgroup
\end{document}